\documentclass{aa}  

\usepackage{graphicx}
\usepackage{txfonts}
\usepackage{hyperref}
\usepackage{subcaption}
\usepackage{amsmath}
\usepackage{svg}
\usepackage{multirow}
\usepackage{float}
\usepackage{orcidlink}
\usepackage[bottom,symbol]{footmisc}

\begin{document} 

   \title{Multi-phase HI clouds in the Small Magellanic Cloud halo}

   \author{F. Buckland-Willis \inst{1,2,\thanks{Corresponding author: francesbw.astronomy@outlook.com}} \orcidlink{0000-0003-4213-8094}
          \and M. A. Miville-Desch\^enes \inst{1,2} \orcidlink{0000-0002-7351-6062}
          \and A. Marchal \inst{3} \orcidlink{0000-0002-5501-232X}
          \and J. R. Dawson \inst{4,5} \orcidlink{0000-0003-0235-3347}
          \and H. D\'enes \inst{6} \orcidlink{0000-0002-9214-8613}
          \and E. M. Di Teodoro \inst{7} \orcidlink{0000-0003-4019-0673}
          \and J. M. Dickey \inst{8} \orcidlink{0000-0002-6300-7459}
          \and S. J. Gibson \inst{9} \orcidlink{0000-0002-1495-760X}
          \and I. P. Kemp \inst{10,11} \orcidlink{0000-0002-6637-9987}
          \and C. Lynn \inst{3} \orcidlink{0000-0001-6846-5347}
          \and Y. K. Ma \inst{3} \orcidlink{0000-0003-0742-2006}
          \and N. M. McClure-Griffiths \inst{3} \orcidlink{0000-0003-2730-957X}
          \and C. E. Murray \inst{12,13} \orcidlink{0000-0002-7743-8129}
          \and N. M. Pingel \inst{14} \orcidlink{0000-0001-9504-7386}
          \and S. Stanimirovi\'c \inst{14} \orcidlink{0000-0002-3418-7817}
          \and J. Th. Van Loon \inst{15} \orcidlink{0000-0002-1272-3017}
          }

   \institute{Laboratoire de Physique de l’Ecole Normale Sup\'erieure, ENS, Universit\'e PSL, CNRS, Sorbonne Universit\'e, Universit\'e de Paris, 24 rue Lhomond, 75005 Paris Cedex 05, France 
   \and AIM, CEA, CNRS, Universit\'e Paris-Saclay, Universit\'e Paris Diderot, Sorbonne Paris Cit\'e, F-91191 Gif-sur-Yvette, France
   \and Research School of Astronomy and Astrophysics, The Australian National University, Canberra, ACT 2611, Australia
   \and School of Mathematical and Physical Sciences and Astrophysics and Space Technologies Research Centre, Macquarie University, 2109, NSW, Australia
   \and Australia Telescope National Facility, CSIRO Space and Astronomy, PO Box 76, Epping NSW 1710, Australia
   \and School of Physical Sciences and Nanotechnology, Yachay Tech University, Hacienda San Jos\'e S/N, 100119, Urcuqu\'i, Ecuador
   \and Dipartimento di Fisica e Astronomia, Universit\`a degli Studi di Firenze, I-50019 Sesto Fiorentino, Italy
   \and School of Natural Sciences, Private Bag 37, University of Tasmania, Hobart, TAS, 7001, Australia
   \and Department of Physics \& Astronomy, Western Kentucky University, 1906 College Heights Blvd., Bowling Green, KY 42101, USA
   \and International Centre for Radio Astronomy Research (ICRAR), Curtin University, Bentley, 6102, WA, Australia
   \and CSIRO Space and Astronomy, 26 Dick Perry Avenue, Kensington, 6151, WA, Australia
   \and Department of Physics \& Astronomy, Johns Hopkins University, 3400 N. Charles Street, Baltimore, MD 21218, USA
   \and Space Telescope Science Institute, 3700 San Martin Drive, Baltimore, MD 21218, USA
   \and Department of Astronomy, University of Wisconsin, Madison, WI 53706-15821, USA
   \and Lennard-Jones Laboratories, Keele University, ST5 5BG, UK
   \\}

   \date{Received 17 October 2024; accepted 29 November 2024}

 
  \abstract
   {The Galactic ASKAP collaboration (GASKAP) is undertaking an HI emission survey of the 21cm line to map the Magellanic system and the Galactic plane with the Australian Square Kilometre Array Pathfinder (ASKAP). One of the first areas observed in the Pilot Phase I of the survey was the Small Magellanic Cloud (SMC). Previous surveys of the SMC have uncovered new structures in the periphery of the SMC, along relatively low column density lines of sight.}
   {In this work we aimed to uncover the phase distribution of three distinct structures in the periphery of the SMC. This work will add to the constraints we have on the existence and survival of the cold neutral medium (CNM) in the SMC. }
   {We used ROHSA, a Gaussian decomposition algorithm, to model the emission across each cloud and classify the HI emission into their respective phases based on the linewidths of the fitted Gaussians. We created maps of velocity and column density of each phase of the HI across these three clouds. We measured the HI mass and CNM number density for each cloud. We also compared the HI results across the different phases with other gas tracers.}
   {We find that in two clouds, the ends of each cloud are almost completely CNM dominated. Analysis of these two clouds indicates they are experiencing a compressive force from the direction of the SMC main body. In the third cloud we find a uniform CNM distribution along one wall of what is likely a supershell structure. Comparison with previous measurements of CO clumps in two of the clouds show the CO and HI are co-moving within a few $\text{km s}^{-1}$ in regions of high HI column density, particularly when considering just the CNM.}
   {}


   \maketitle
%

\section{Introduction}
\label{sec:intro}

The interstellar medium (ISM) consists of many different components, of which one of the most abundant is neutral hydrogen (HI). The study of neutral hydrogen helps us to understand the distribution and conditions needed to facilitate star formation. Understanding the ways in which HI cools and condenses to become molecular hydrogen is a key part of the star formation process. 

One factor that affects the cooling ability of the ISM is the metallicity. Understanding how metallicity affects the distribution of the neutral hydrogen, requires studying the HI in different environments. The Magellanic Clouds provide us with an excellent, nearby laboratory in which to study HI in low-metallicity environments. The Large Magellanic Cloud (LMC) has a characteristic metallicity of 0.5 Z$_
{\odot}$ \citep{rolleston2002PresentdayChemical} and the Small Magellanic Cloud (SMC) has one of 0.2 Z$_
{\odot}$ \citep{russell1992AbundancesHeavy}. Both clouds provide low-metallicity environments, with significant HI gas reservoirs that can be well-resolved in observations. With the advent of high-resolution, high-sensitivity interferometers such as the upcoming Square Kilometre Array (SKA) \citep{mcclure-griffiths2015GalacticMagellanic}, galaxies such as the Magellanic Clouds will be imaged at the same physical scales as the Milky Way had been previously, allowing for a close comparison of a broad range of Galactic environments.

The SMC is an irregular galaxy, whose 3D structure is quite complex. Previous HI studies of the SMC have shown that the neutral gas is very dynamically complex, especially along lines of sight in the central areas of the galaxy. The neutral gas distribution has been shaped by the interaction of the SMC with the LMC, simulated in \citet{diaz2011ConstrainingOrbital} and \citet{besla2012RoleDwarf}, as well as star formation activity in the galaxy. Notably, in the HI survey conducted with a combination of Australian Telescope Compact Array (ATCA) and Murriyang (Parkes) observations \citep{staveley-smith1997HIAperture} just over 500 shells were catalogued, indicating that turbulence, star formation and their resulting winds, and supernovae are changing the HI landscape. These processes inject energy into the medium which can compress the neutral hydrogen, forming the dense gas needed to form the next generation of stars \citep{inutsuka2015FormationDestruction,dawson2013SupershellMolecular}. Disruption from phenomena occurring within the galaxy and the tidal forces from the larger Magellanic system interaction has made the SMC a very disturbed and dynamic system. Stellar studies have shown that the galaxy's stellar population is extended by 1-10 kpc along the line of sight \citep{muraveva2018VMCSurvey}. A recent study \citep{murray2024GalacticEclipse} comparing the HI and stellar population of the galaxy suggests that there are two main sections of the galaxy positioned at differing distances along the line of sight. This elongation, coupled with the fact that HI spectral data give only velocity information along the line of the sight, makes it particularly difficult to reliably determine distances to any discrete structures in the galaxy. 

Despite the challenges in studying the HI distribution of the SMC, its low metallicity provides a valuable environment with different conditions to that of the Milky Way. Because the main cooling mechanisms of the HI are less efficient at lower metallicities \citep{wolfire1995NeutralAtomic}, the distributions of the warm neutral medium (WNM) compared to the cold neutral medium (CNM) through the SMC should differ when compared to the HI under Galactic conditions. These different phases, WNM and CNM, are defined by their typical temperatures in the Milky Way in \citet{wolfire1995NeutralAtomic} and \citet{wolfire2003NeutralAtomic}, where WNM (T $\sim 8000$ K) and CNM (T $\sim 100$ K). Another phase of the HI, the unstable neutral medium (UNM) is often seen in studies \citep{mcclure-griffiths2023AtomicHydrogen}. This phase is HI which lies in the thermally unstable range of pressures and temperatures, and it is a transitory state of HI that will evolve into either stable phase of the HI (CNM or WNM). The exact temperatures and pressures that define all of these phases depend on the ISM conditions, as detailed in \citet{wolfire1995NeutralAtomic} and \citet{wolfire2003NeutralAtomic}.

The CNM distribution in the SMC has been studied previously through absorption surveys, both targeted and un-targeted. Across three previous absorption studies \citep{dickey2000ColdAtomic,jameson2019ATCASurvey,dempsey2022GASKAPHIPilot} the reported mean CNM fraction for the SMC has ranged from $0.07-0.2$. \citet{dempsey2022GASKAPHIPilot} reported CNM fraction values for individual lines of sight, which show that there are a few lines of sight with a CNM fraction close to $0.5$, but the majority are below $0.2$. CNM is found to populate discrete filamentary structures in the ISM \citep{mcclure-griffiths2023AtomicHydrogen} which cannot be well-characterised by absorption studies alone. This is because, while absorption studies are very effective at identifying CNM, especially when the gas is very cold and easily absorbs, the detections of the CNM are limited by the background source density. Therefore absorption studies cannot provide a full map of the CNM across any specific field. 

New developments in Gaussian decomposition and other HI analysis methods have helped to map the CNM fraction across large fields. In particular, Regularized Optimization for Hyper-Spectral Analysis (ROHSA), first introduced in \citet{marchal2019ROHSARegularized}, has since been used to map the different HI phases across a number of Galactic structures \citep{marchal2021ResolvingFormation,taank2022MappingThermal,vujeva2023MappingMultiphase}. The ROHSA algorithm takes advantage of the expected spatial coherence of the HI signal and enforces similar solutions between neighbouring sightlines. This allows the distribution of the different phases of the HI across the field to be tracked directly from the emission spectra. Of course, it is vital to have the appropriate spectral resolution in order to resolve the CNM signals that typically have linewidths $< 3\text{ km s}^{-1}$. Another method, using Fourier transforms of the spectral axis of HI data, demonstrated in \citet{marchal2024MappingLower}, estimates a lower limit for the CNM fraction with great speed across large fields of view. 

Analysis of data with these tools will help refine techniques in preparation for HI surveys currently observing with the Australian Square Kilometre Array Pathfinder (ASKAP) telescope and the data that will eventually come from the SKA-MID (Square Kilometre Array MID) telescope. These telescopes provide datasets with improved angular resolution and sensitivity, as well as adequate velocity resolution to uncover the distribution of the CNM within and beyond the Milky Way. In this work we look at three discrete HI clouds within the periphery of the SMC, first identified in \citet{mcclure-griffiths2018ColdGas}, with data from the Pilot phase of the Galactic ASKAP HI (GASKAP-HI) survey and we analysed them using the ROHSA algorithm.

In Section \ref{sec:data} we outline the data we analysed for this work and the noise estimation. In Section \ref{sec:methods} we detail the modelling of the HI data using ROHSA, describing the parameters used in the fitting process and the associated uncertainties. In Section \ref{sec:results} we describe the results from the fitting process and the main trends we see across the three clouds. In Section \ref{sec:discussion} we discuss the HI in relation to other tracers CO and H$\alpha$, as well as each of its phases, and we discuss formation scenarios for each of these clouds.

\section{Data}
\label{sec:data}
\subsection{GASKAP collaboration data}
\label{subsec:gaskapdata}

The data in this work comes from the GASKAP-HI survey Pilot Phase I observations of the SMC. A total of 20.9hr of integration was undertaken over two observing blocks in December 2019. The data was reduced with a joint-imaging pipeline during which it was combined with single dish observations from the Murriyang telescope, the details are contained in the data release paper \citep{pingel2022GASKAPHIPilot}. After processing, the hyperspectral data (in the form of a position-position-velocity, PPV, cube) for the SMC field was imaged at an angular resolution of 30" and spectral resolution of 0.98 $\text{ km s}^{-1}$ (smoothed from the native resolution of $\approx 0.3 \text{ km s}^{-1}$). The field of view of the datacube is $\approx 5\times 5 \text{ deg}^2$, which nicely encompasses the majority of the HI distribution of the SMC, including the lower column-density areas with gas that leads to the Bridge and the Counter-Arm. The velocities of the SMC datacube range from $40.0 \text{ km s}^{-1} - 253.9 \text{ km s}^{-1}$, capturing the full dynamic range of the SMC. 

\subsection{Noise estimation}
\label{subsec:noiseestimation}

For this dataset, the median noise level measured across the whole cube was reported as 1.1 K in \citet{pingel2022GASKAPHIPilot}. As we consider specific regions of the SMC as opposed to the SMC as a whole, it is necessary to have a local estimation of the noise. Thus, in this work, we take an empirical approach to obtain the noise values across the datacube. This method requires identifying channels in the datacube that are 'emission-free', from which to measure the rms noise at each pixel. The velocity range of the datacube encompasses the entirety of the expected SMC emission and extra channels either side. Through visual inspection, channels where $v<62\text{ km s}^{-1}$ and $v>235\text{ km s}^{-1}$ were determined to be free of SMC emission, allowing 43 of 220 channels (19.5\%) to be used to calculate the rms noise level of each pixel. Taking the rms of the values across the 43 channels resulted in a median noise value of 1.6 K, with the values increasing towards the edges of datacube as the beam response decreases. The noise in the HI line increases with the strength of the emission, so we calculate the noise along each line of sight as is done in \citet{boothroyd2011AccurateGalactic}:
\begin{equation}
\centering
    \sigma(v)= \sigma_0\left(\frac{T_\text{sys}+ T_\text{B}(v)}{T_\text{sys}}\right),
\label{eq:sysnoise}
\end{equation}
where $\sigma_0$ is the rms noise level of an individual pixel. We use the provided value of $T_\text{sys}=55$ K from \citet{pingel2022GASKAPHIPilot}. With this added dimension, we obtain a 3D map of the noise throughout the entire datacube.

\section{Methods}
\label{sec:methods}

\subsection{Field selection}
\label{subsec:dataselection}
The fields analysed in this work were chosen to encompass three HI clouds first identified in \citet{mcclure-griffiths2018ColdGas} as outflows from massive star formation within the SMC. These clouds were chosen because their emission is very strong in the densest regions of the clouds, with column densities of the order of $10^{20}$ cm$^{-2}$. Additionally, while the distribution of CNM in the main body of the SMC can be studied in absorption, these clouds have compact structures and were not probed by any background sources in the latest absorption study \citep{dempsey2022GASKAPHIPilot}, due to the intrinsic limit imposed by the background source density. Thus, this work provides an opportunity to analyse the phase distribution just from the emission of HI clouds with high signal to noise ratios. 

\begin{figure*}[h]
    \includegraphics[width=\textwidth]{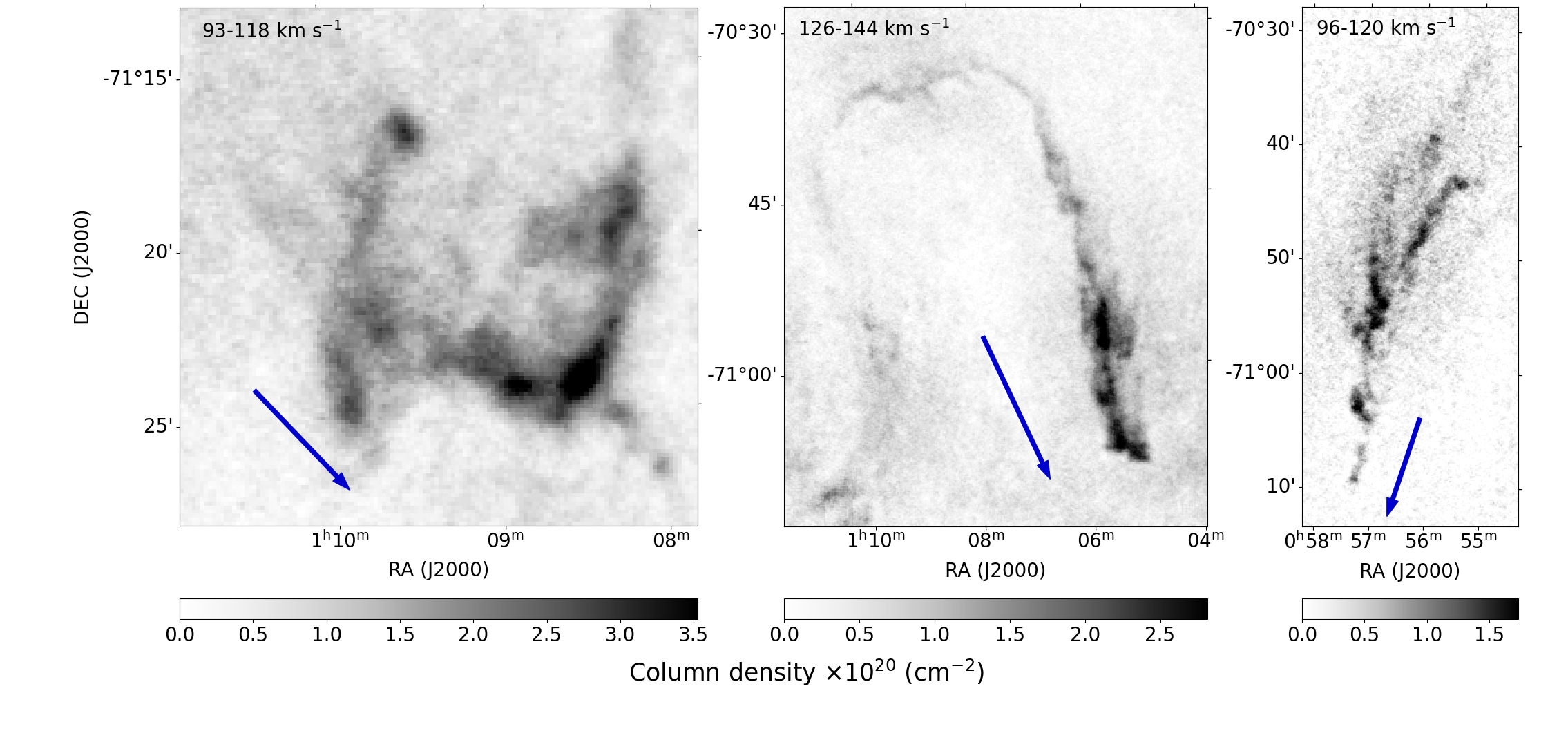}
    \caption{Column density images of all three clouds (Left: Alpha, Middle: Hook, Right: Gamma) produced by integrating each field over the velocity range indicated in each panel. The blue arrows indicate the direction of the dynamical centre of the SMC derived in \citet{diteodoro2019DynamicsSmall}.}
    \label{fig:all3columndensity}
\end{figure*}

In Figure \ref{fig:all3columndensity} we show the total column density for each cloud, which we name Alpha, Hook, and Gamma, respectively, over the velocity range they span. Additionally, we also indicate the direction of the SMC dynamical centre as reported in \citet{diteodoro2019DynamicsSmall}. Each cloud has a different morphology, with the Alpha cloud being the smallest of the three, which assuming a distance of 63 kpc \citep{diteodoro2019DynamicsSmall} measuring 240 pc across the longest diagonal (north-east to south-west), with an irregular morphology. The Hook cloud is primarily composed of the strong ridge along the west side forming a long filament that spans 610 pc. It also continues looping around the north and down the east of the field, albeit with lower column densities. The Gamma cloud has a strong core filament, and appears to be broken into multiple clumps. It covers a distance of 560 pc, similar to the Hook cloud, but with a clumpier appearance.

\subsection{Gaussian decomposition with ROHSA}
\label{subsec:gaussdecomp}

\subsubsection{Defining a best-fit solution}
\label{subsubsec:bestfit}
We used the ROHSA algorithm developed in \citet{marchal2019ROHSARegularized} to spectrally decompose each of the three clouds. The ROHSA tool allows us to take advantage of the expected spatial coherency of the HI signal, that the decomposition solution along a line of sight is similar to that of its neighbours. This approach helps tackle the degeneracy problem that can muddy HI decomposition efforts along complicated lines of sight with multiple components with differing spreads and centres. The degree to which this coherency is enforced is controlled by 4 hyperparameters, detailed in Section \ref{subsubsec:fitting}, and it is the fine-tuning of these hyperparameters that will have the most impact on the final solution. As part of this work, an exploration of the 4-D hyperparameter space was undertaken to determine the best solution. We define a best solution as one whose residuals are normally distributed and achieves a reduced chi-squared value close to 1 with the fewest number of components.

\subsubsection{Separating cloud and main body emission}
\label{subsubsec:separatespec}
\begin{figure}[h]
    \centering
    \includegraphics[width=0.5\textwidth]{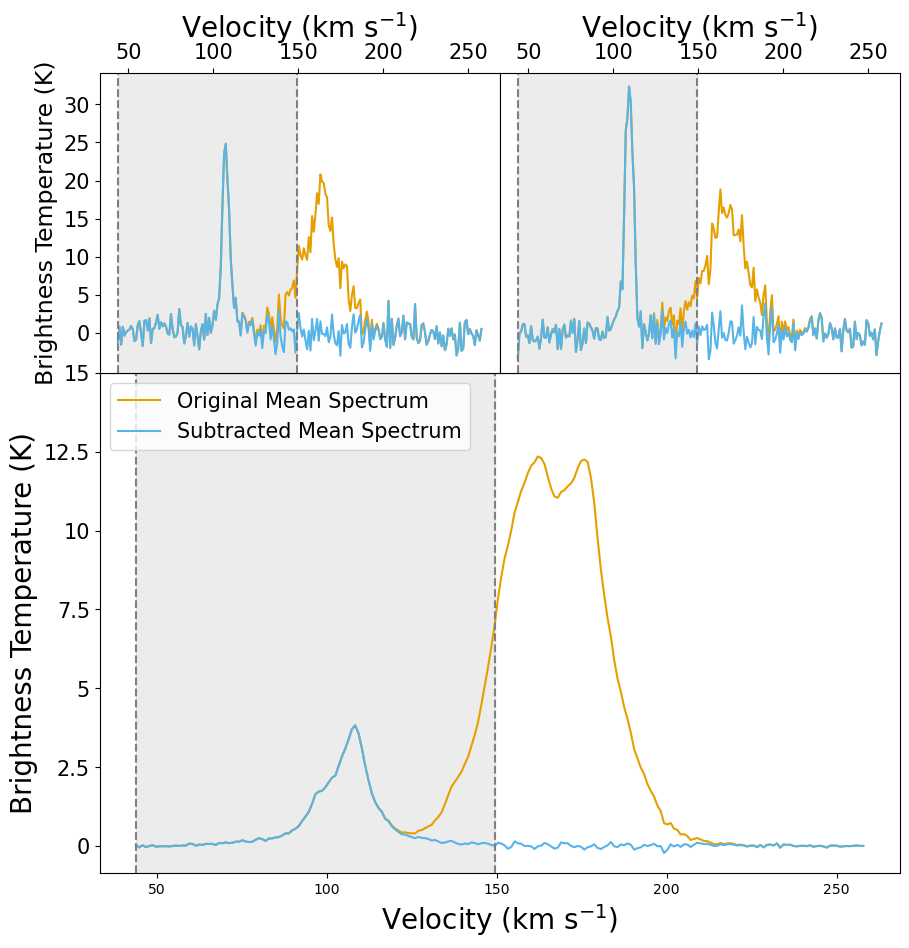}
    \caption{Example spectra from the Alpha cloud field. Bottom: Mean spectrum of the field before (orange) and after (blue) subtraction of the main body emission signal. Top: spectra of two separate sightlines with the same colours indicating the spectra before and after subtraction. The grey windows show the spectral range retained for the fitting described in Section \ref{subsubsec:fitting}.}
    \label{fig:alphasubspec}
\end{figure}

While the signal from the three target clouds is strong in the brightest areas, the emission from the main body of the SMC dominates the mean spectrum over each full field containing the clouds, shown in Figure \ref{fig:alphasubspec}.  In preliminary attempts to fit the full spectra, 8-9 components were used, when only 2 components are likely necessary to model the emission from the clouds themselves. If there are not enough components to fully model the main body and cloud emission across the whole field then the weakest emission channels will be un-fit. However, adding additional components to the model can increase the degeneracy of the solution and increase the computational cost. Since the emission from the main body of the SMC is of little interest to us in this work, we opted to isolate the cloud emission to simplify the fitting efforts.

To overcome this problem, we initially took a subset of the spectrum over the velocity range of interest and proceeded to fit just this section. Although the targeted clouds are removed from the main body emission of the SMC, there are some wide (WNM) components from the main body whose tails bleed into the velocity range of interest for these clouds. If we do nothing but take a subset of the data over the velocities of interest, we risk having these tails disrupt ROHSA's fitting efforts. If we allow ROHSA to fit this unrefined data, it will not find the true signal that this tail belongs to, as the central velocity now lies outside the velocity range it is allowed to consider, leading to one of two scenarios. Either the tail is not fit at all by the solution, increasing the residual, or; the tail is fit by a relatively thin component that will likely still leave a non-noise residual and artificially inflate the CNM fraction found by the solution. Neither scenario is preferable. To combat this, we do a preliminary fit to the entire spectrum with low amounts of regularisation to identify the bulk main-body emission of the SMC. This allows us to remove the redundant emission before segmenting the spectra to the velocity range that we are interested in. Once the main body emission has been removed we discard this part of the spectrum, focusing on a subset of channels over which the feature of interest exists. In this removal stage we are not interested in individual spectral components or concerned with the degeneracy of solution, we merely aim to recreate the shape of the signal without the noise, to remove it. Now we have a reduced dataset that requires fewer components, where a stable solution is more achievable. Figure \ref{fig:alphasubspec} shows the mean spectrum before and after subtraction of the main body emission as well as the spectra from two non-neighbouring sightlines. The grey windows in Figure \ref{fig:alphasubspec} show the reduced spectral range used for fitting in the next Section.

\subsubsection{Fitting processed spectra}
\label{subsubsec:fitting}

After removal of the main-body emission, the final step is to fit the features of interest. At this point, we want to optimise the hyperparameters used by ROHSA to constrain the solution. Additionally, we keep in mind from our definition of best fit solution in Section \ref{subsubsec:bestfit}, that we want to use as few components as possible. The cost function of ROHSA contains a regularisation term that is set by four hyperparameters. The first three hyperparameters act to regularise the elements of each Gaussian, that is. the amplitude ($A$), the central velocity ($\mu$) and the velocity dispersion ($\sigma$) within a 1 pixel radius. The fourth hyperparameter regularises the value of $\sigma$ across the entire field, as it corresponds to the phase properties of that component and should not vary to a significant degree. These hyperparameters are incorporated into the cost function ($J(\theta, v)$) as such, from \citet{marchal2019ROHSARegularized}:

\begin{align}
\centering
    J(\theta, v) & = L\left(\theta, v\right) + \frac{1}{2}\sum^N_{n=1}\lambda_A||DA_n||_2^2 + \lambda_\mu||D\mu_n||_2^2 + \lambda_\sigma||D\sigma_n||_2^2 \nonumber \\
    & + \lambda_\sigma'||\sigma_n - m_n||_2^2,
\label{eq:ROHSAcost}
\end{align}

where $\theta$ represents the positional axes and $v$ is the velocity axis, both in the hyperspectral data. $N$ is the number of Gaussians. $\lambda_a$, $\lambda_{\mu}$, $\lambda_{\sigma}$, and $\lambda_{\sigma}'$ are the hyperparameters of ROHSA that control the spatial variance of the fitted parameters $A$ (amplitude), $\mu$ (central velocity), and $\sigma$ (velocity dispersion). $m$ represents a mean value for $\mu$ of each $N$ Gaussians. For further details, refer to \citet{marchal2019ROHSARegularized}.

\begin{figure}[h]
    \centering
    \includegraphics[width=0.5\textwidth]{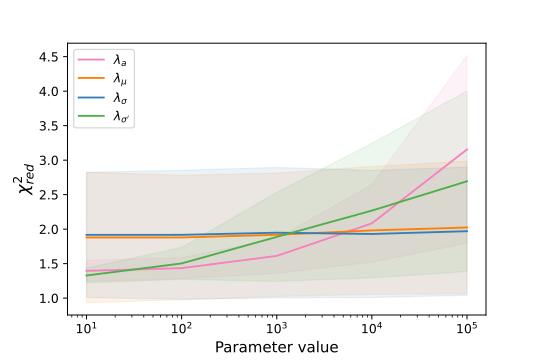}
    \caption{Change of average $\chi_\text{red}^2$ value as all hyperparameters are varied for the Gamma cloud field. Shaded regions show the standard deviation around the mean.}
    \label{fig:hypechange}
\end{figure}

We explored the parameter space in orders of magnitude for each hyperparameter for the Gamma cloud field. Figure \ref{fig:hypechange} shows how the reduced chi-squared value ($\chi_\text{red}^2$) for the Gamma cloud field changes as a function of the hyperparameter value. Evidently, the hyperparameters that most affected the $\chi_\text{red}^2$ were $\lambda_A$ and $\lambda_{\sigma}'$. On the other hand, $\lambda_v$ and $\lambda_{\sigma}$ did not change the $\chi_\text{red}^2$ very much.

This field was explored extensively by initialising ROHSA with each hyperparameter varying over 5 orders of magnitude. This amounts to $5^4 = 625$ runs over a single field. This is very computationally expensive way to narrow down the correct hyperparameters for a single field. So for the remaining 2 fields, a more scaled-back approach was taken. We only explored a hyperparameter space one order of magnitude either side of the best fit hyperparameters that were obtained for the Gamma cloud field (indicated in Table \ref{tab:cloudfits}). 

This smaller exploration resulted in a different outcome, that the hyperparameter values did not affect the average $\chi_\text{red}^2$ values by an appreciable amount. We found it was difficult to ascertain, from just the average value of the $\chi_\text{red}^2$, what the best fit solution was. Thus, we have the stipulation in our definition that the residuals must be normally distributed. If the field is under fit the distribution would be positively skewed, and on the other hand if it was over fit, the distribution would be negatively skewed.

The number of Gaussians fit to each field were selected through a similar exploration of the parameter space. From visual inspection of the spectra in each cloud, it was estimated that at least two Gaussian components were required to capture the full signal from each cloud. From this starting point, we obtained ROHSA solutions for each cloud ranging from 2-4 Gaussians. We assessed the fits of differing amount of Gaussians by inspecting the map of reduced chi-squared values calculated for each field for each line of sight. In a well-fit solution, this map should resemble a grid of random noise. We found in all three fields that we were able to achieve this noise-like reduced chi-squared map with three Gaussians as a minimum. In all fields, using only two Gaussians left parts of spectra unfit and there was clear structure in the reduced chi-squared maps. We achieved very slight reductions in the mean reduced chi-squared values with four Gaussians, however the reduced chi-squared maps did not appear to change significantly. It is important to cease adding additional Gaussian components to spectra once the best-fit criteria are satisfied. It is also valuable to note that ROHSA allows Gaussian components to have an amplitude of zero if that solution minimises the cost function (Equation \ref{eq:ROHSAcost}). So, to say that a three Gaussian solution is fit to the field does not denote that all three Gaussians are actively contributing to the solution for every spectrum in the field. It should be thought of as the maximum number of Gaussians fit to any one spectrum in the field. So after inspecting the distribution of residuals and verifying that they are centred around zero we selected the three Gaussian solutions for each of the field. These solutions are discussed in Section \ref{subsec:bestfit}.

\subsection{Error estimation}
\label{subsec:errorestimate}
Errors are not obtained directly from the ROHSA output, so we follow the method outlined in \citet{taank2022MappingThermal} in section 4.2.2. This involves three distinct methods of estimating error and taking their quadratic sum as the total uncertainty. These methods all test the stability of the solution. The first method involves taking the solution and adding random noise onto it, the levels of which are determined by the noise calculated in section \ref{subsec:noiseestimation}. The ROHSA algorithm is then run on these noise realisations with the same hyperparameter values as the original solution. The second method takes one of the noise realisations and runs ROHSA with hyperparameter values that vary by up to 10\% around the values used to achieve the original solution. The hyperparameter values for each of these runs were drawn from a uniform distribution within 10\% of the original hyperparameter values. The third method takes the same noise realisation used in the previous method and varies the initial guess ROHSA uses to fit the top-level solution. The spread of fitted values for each component of each Gaussian fitted in the original provides a reasonable range from which to select the initial guesses, as it is sufficiently variable, without being random. The range was thus taken as the FWHM of the distribution of each Gaussian component around its mean. The initial guesses for each of these runs were drawn from a uniform distribution within the aforementioned range. For each of these three methods, 100 runs were completed, totalling 300 runs with ROHSA. To assess the uncertainty from each method, the standard deviation in each Gaussian parameter from the 100 runs was calculated. To then obtain the total uncertainty ($\sigma_\text{tot}$) from these three methods, we take the square root of the quadratic sum as per Equation \ref{eq:error},

\begin{equation}
\centering
    \sigma_\text{tot} = \sqrt{\sigma_\text{rn}^2 + \sigma_\text{hp}^2 +    \sigma_\text{ig}^2},
\label{eq:error}
\end{equation}
where $\sigma_\text{rn}$ is the uncertainty from the first method with random noise, $\sigma_\text{hp}$ is the uncertainty from the second method with hyperparameter variation, and $\sigma_\text{ig}$ is the uncertainty from the third method with initial guess variation. In \citet{taank2022MappingThermal}, they found that the initial guess variation was the largest source of error out of the 3 for their 11 Gaussian fit. Once we calculated our errors from all 3 sources, we found the error from the noise realisations is consistently the largest source of error. The ratio of the contribution from noise realisations, hyperparameter change and initial spectrum for the Alpha field is 48:25:27, for the Hook field is 44:14:42, and for the Gamma field is 55:32:13. In all 3 fields the noise uncertainty contributes most to the overall uncertainty, whereas the other two sources have differing effects on each of the fields. We find that varying the initial guess does not dominate the uncertainty in these fields as it does in \citet{taank2022MappingThermal}, due to the relative simplicity of our 3 Gaussian models compared to their 11 Gaussian model. Degenerate solutions become more abundant when the number of Gaussians is increased and thus solutions with fewer Gaussians will be more robust against changes to the initial guess. 
\section{Results}
\label{sec:results}

\begin{table*}
\small
\caption{Fitting parameters and results for each cloud field in this work.}             
\label{tab:cloudfits}      
\centering          
\begin{tabular}{c c c c c c c c c c c c c c c c }     
\hline\hline        
\multirow{3}{*}{Cloud} & \multirow{3}{*}{n} & \multirow{3}{*}{$\lambda_A$} & \multirow{3}{*}{$\lambda_\mu$} & \multirow{3}{*}{$\lambda_\sigma$} & \multirow{3}{*}{$\lambda_\sigma'$} & \multirow{3}{*}{$\overline{\chi^2_{\text{red}}}$} & \multicolumn{3}{c}{Component 1} & \multicolumn{3}{c}{Component 2} & \multicolumn{3}{c}{Component 3}\\ 
 & & & & & & & $A$ & $v$ & $\sigma_v$ & $A$ & $v$ & $\sigma_v$ & $A$ & $v$ & $\sigma_v$\\
 & & & & & & & (K) & (km s$^{-1}$) & (km s$^{-1}$) & (K) & (km s$^{-1}$) & (km s$^{-1}$) & (K) & (km s$^{-1}$) & (km s$^{-1}$) \\
\hline
\vspace{1\baselineskip} \\
   Alpha & 3 & 1 & 1 & 1 & 100 & 1.17 & 2.35 & 108 & 7.77 & 6.51 & 104 & 1.97 & 0.43 & 79.94 & 9.35\\  
   Hook & 3 & 1 & 10 & 10 & 100 & 1.19 & 2.87 & 135 & 3.99 & 0.71 & 124 & 6.17 & 2.84 & 132 & 1.32\\
   Gamma & 3 & 10 & 10 & 10 & 10 & 1.12 & 0.99 & 102 & 6.48 & 1.81 & 113 & 1.29 & 2.65 & 107 & 1.48\\
\hline                  
\end{tabular}
\tablefoot{Column 1 gives the number of components fit to the field, columns 2-5 give the values of each hyperparameter for the fit with ROHSA, column 6 gives the reduced chi-squared value for each fit. The remaining columns show for each Gaussian component that was fit to the fields, the mean values of the amplitude ($A$), central velocity ($\mu$), and velocity dispersion ($\sigma$).}
\end{table*}

\subsection{Best-fit models} 
\label{subsec:bestfit}
After finding solutions that satisfy our previous definition of a best fit, we list the best fit parameters in Table \ref{tab:cloudfits}. We classify the components into the different HI phases based on the mean maximum kinetic temperature ($\left<T_k\right>$) measured from the mean Gaussian dispersion ($\left<\sigma\right>$) for each component. The maximum kinetic temperature is the temperature of the gas if there was no turbulent broadening of the HI signal. In reality there is some contribution from turbulent broadening, so the maximum kinetic temperature is an upper limit on the true temperature of the gas. The maximum kinetic temperature is derived from the following relation from \citet{draine2011PhysicsInterstellar}:

\begin{equation}
\centering
    \left<T_k\right> = 121\left<\sigma\right>^2.
\label{eq:kintemp}
\end{equation}
Gaussian components with $\left<T_k\right><500$ K are classified as CNM, those with $500 <\left<T_k\right> <5000$ K are classified as UNM, and those with $\left<T_k\right>>5000$ K are classified as WNM, to be consistent with the phase definitions in  \citet{heiles2003MillenniumArecibo}.

\begin{figure*}[h]
\includegraphics[width=\textwidth]{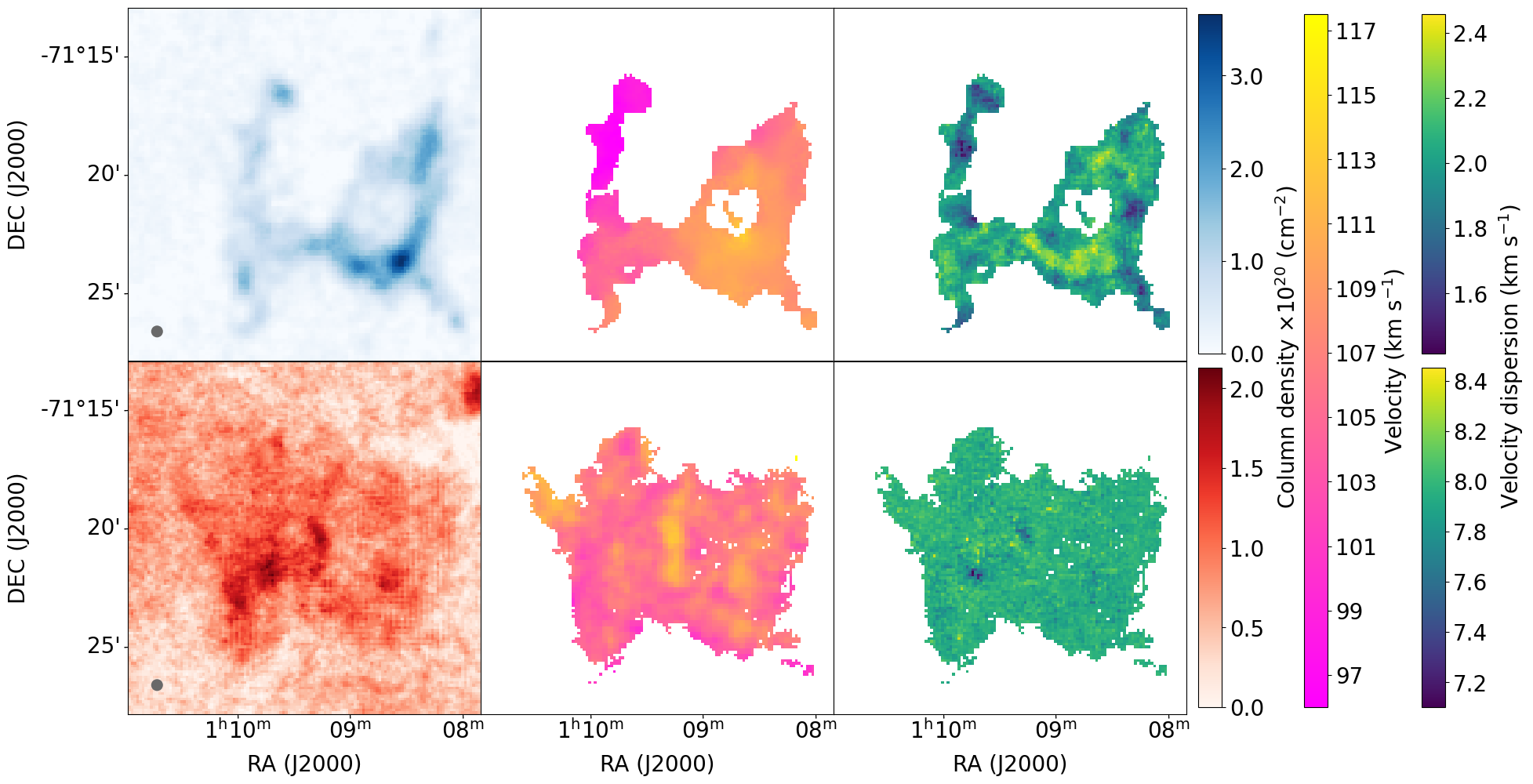}
\caption{Spatial maps of the fitted properties of the Alpha cloud from ROHSA within a $1 \times 10^{20} \text{ cm}^{-2}$ column density contour for each phase. \textit{Top:} Total CNM, \textit{Bottom:} Total WNM. \textit{Left:} Column density ($N_{\text{H}}$), \textit{Centre:} central velocity ($\mu$), \textit{Right:} velocity dispersion ($\sigma$). The grey ellipse in the column density maps indicates the beamsize.}
\label{fig:bigalphafig}
\end{figure*}

For the Alpha cloud we obtained a best fit model with 3 components, with 2 of these components clearly corresponding to the cloud. The CNM component has $\left<T_k\right> = 470$ K and the WNM component has $\left<T_k\right> = 7305$ K. In Figure \ref{fig:bigalphafig} the CNM component traces the strong filamentary structure of the cloud as we see it in the integrated data, whereas the WNM component has a more extended distribution. However, the two components overlap in the velocity axis (seen in Figure \ref{fig:bigalphafig}), showing the components are related to each other in the position and velocity axes. The third unrelated component we obtain for this fit has a large average velocity dispersion from which it would be classified as WNM. On average, its velocity offset is 25-30 km s$^{-1}$ from the cloud, thus it is likely diffuse, low-level WNM emission at the extreme velocity ends of the SMC. Within the high column density areas of the cloud, this unrelated component contributes negligibly to the total column density. For this fit we achieve a column density weighted mean $\chi^2_{red}$ value of 1.17 and we show the column density weighted mean values for each component parameter in Table \ref{tab:cloudfits}.

\begin{figure*}[h]
\includegraphics[width=\textwidth]{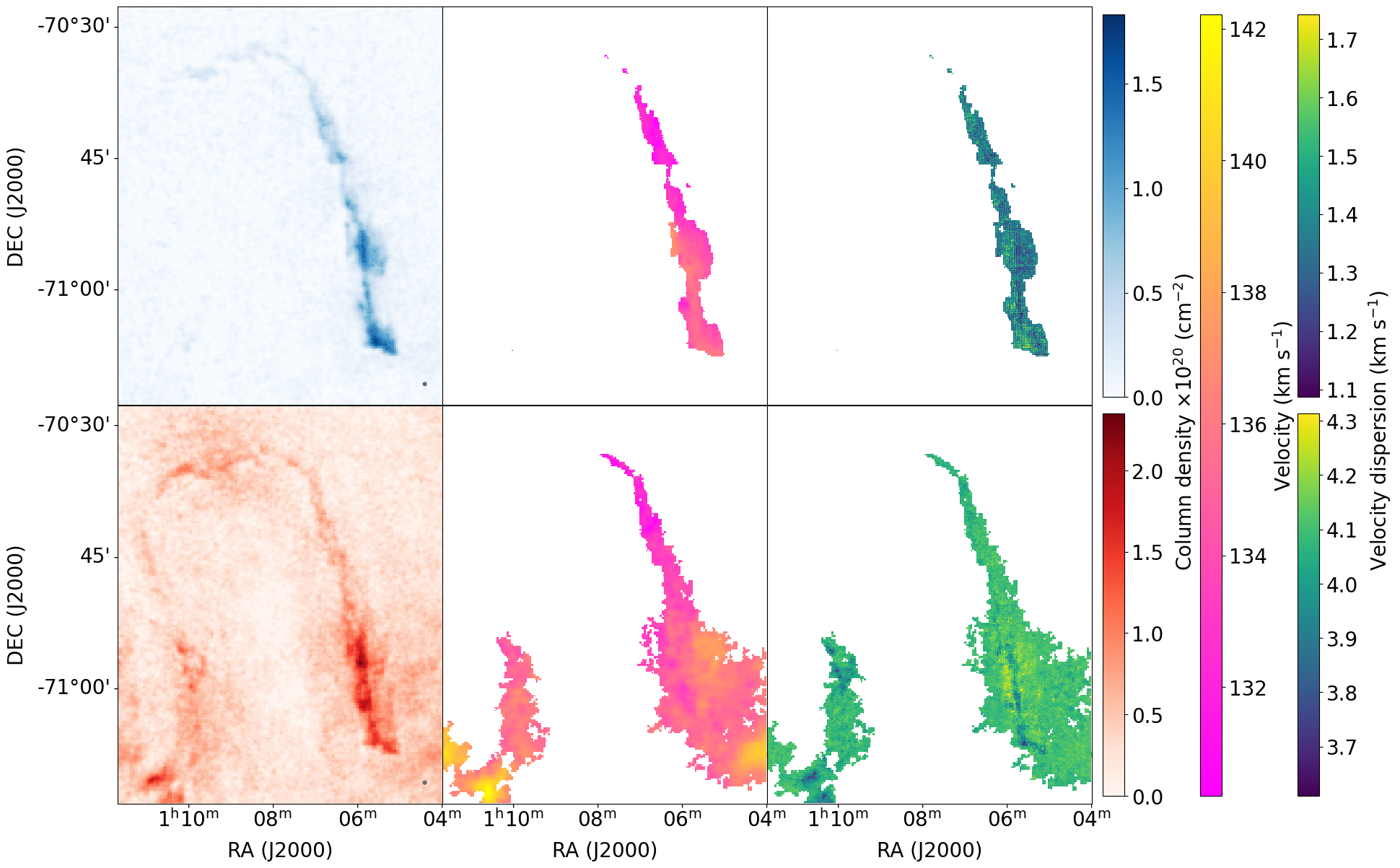}
\caption{Spatial maps of the fitted properties of the Hook cloud from ROHSA within a $5 \times 10^{19} \text{ cm}^{-2}$ column density contour for each phase. The panels are the same as in Figure \ref{fig:bigalphafig}.}
\label{fig:bighookfig}
\end{figure*}

For the Hook cloud we obtained a best fit model again with 3 components, with just 2 of these components clearly corresponding to the cloud. In this case we do not detect significant WNM emission, but rather a cooler UNM component. The CNM component has $\left<T_k\right> = 211$ K and the UNM component has $\left<T_k\right> = 1926$ K. The CNM component traces the strongest part of the Hook shape along the west and into the north, but does not extend into the eastern part of the field. The UNM component envelops the CNM component and loops all the way around the field from west to east. From Figure \ref{fig:bighookfig} we can see that they have the same central velocity for the areas where they overlap on the sky, discussed further in Section \ref{subsec:velgrad}. The 3rd component for this fit is unrelated to the Hook cloud. It does not trace any structure of the cloud, and does not follow the same velocity trend. It is likely a tracing diffuse WNM signal, as in the Alpha cloud. For this fit we achieve a column density weighted mean $\chi^2_{red}$ value of 1.19 and we show the column density weighted mean values for each component parameter in Table \ref{tab:cloudfits}.

\begin{figure*}[t]
\centering
\includegraphics[height=0.85\textheight]{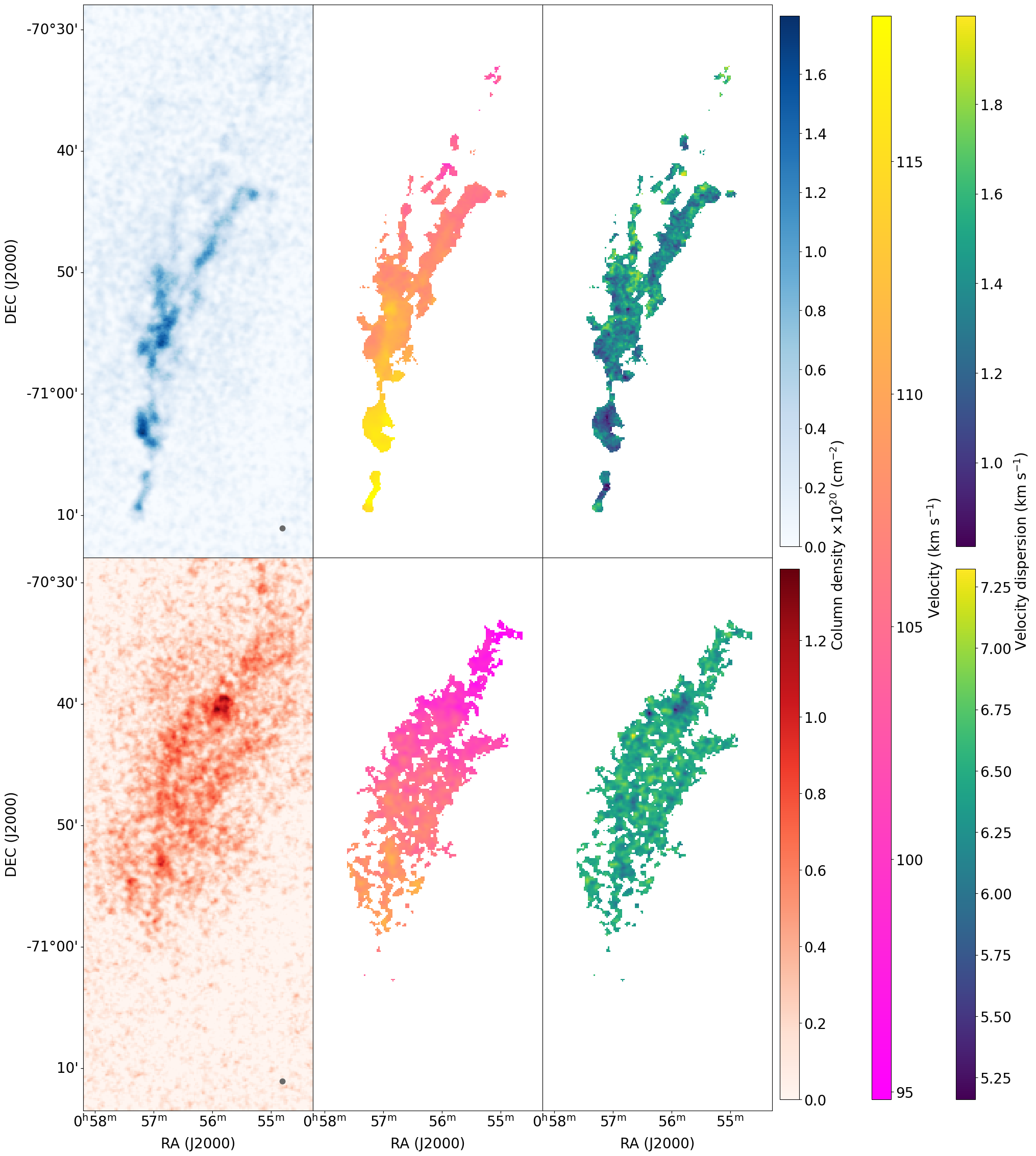}
\caption{Spatial maps of the fitted properties of the Gamma cloud from ROHSA within a $5 \times 10^{19} \text{ cm}^{-2}$ column density contour for each phase. The panels are the same as in Figure \ref{fig:bigalphafig}.}
\label{fig:biggammafig}
\end{figure*}

For the Gamma cloud we obtained a best fit model again with three components, with all three components contributing to the total HI of the cloud. There are two CNM components, with $\left<T_k\right> = 201$ K and $\left<T_k\right> = 265$ K and one WNM component with $\left<T_k\right> = 5081$ K. Like in the Alpha cloud, we see that the CNM and WNM components overlap in velocity space, shown in Figure \ref{fig:biggammafig}. This fit required 2 components to fully capture the CNM distribution as there are strong, narrow signals at offset velocities in parts of this field. Table \ref{tab:cloudfits} shows that they are on average separated by 6 km s$^{-1}$. The advantage of the regularisation conditions that ROHSA imposes is that we can trace how these components evolve individually over the spatial axes, further discussed in Section \ref{subsec:velgrad}. For the purpose of mapping the velocity and velocity dispersion of the total CNM, we take the column density weighted mean of the two components in Figure \ref{fig:biggammafig}. For this fit we achieve a weighted mean $\chi^2_{red}$ value of 1.12 and we show the column density weighted mean values for each component in Table \ref{tab:cloudfits}.

\subsection{Phase distribution}
\label{subsec:phasedist}

\begin{figure*}[h]
\includegraphics[width=\textwidth]{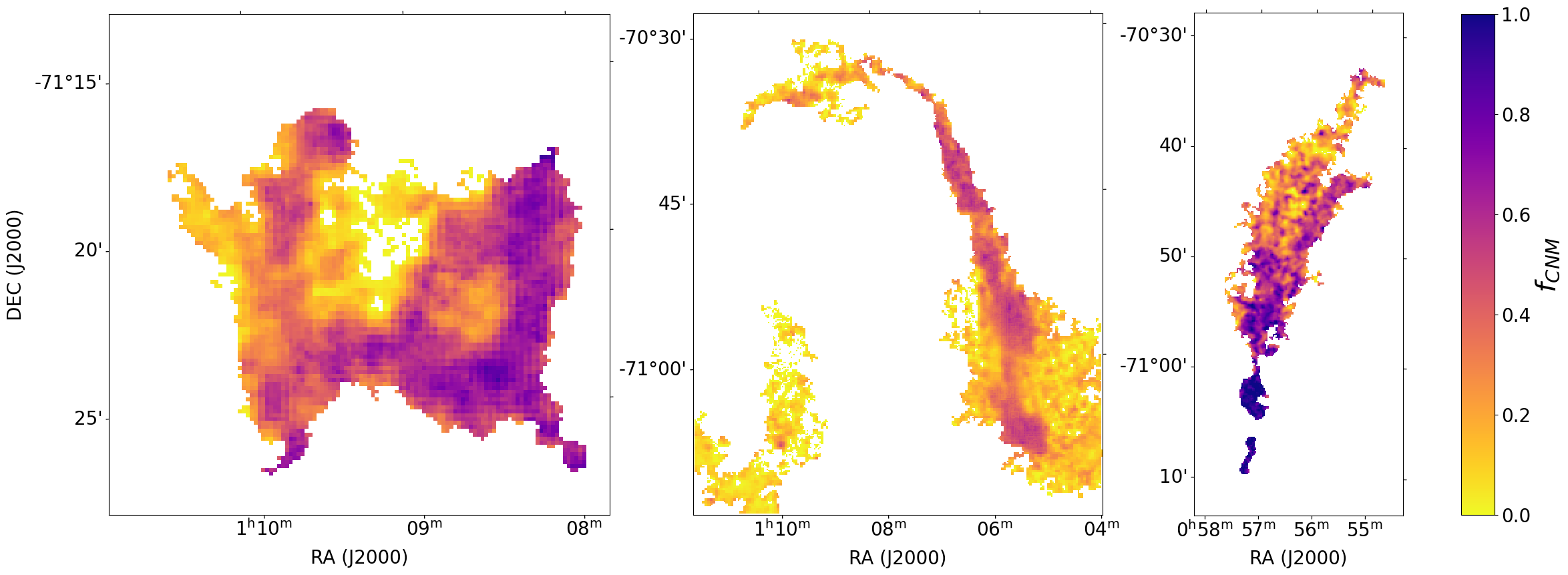}
\caption{CNM fraction ($f_{\text{CNM}}$) of all three clouds (Left: Alpha, Middle: Hook, Right: Gamma) within a $1 \times 10^{20} \text{ cm}^{-2}$ total column density contour for the Alpha cloud and within a $5 \times 10^{19} \text{ cm}^{-2}$ total column density contour for the Hook and Gamma clouds.}
\label{fig:phasedist}
\end{figure*}

The CNM fraction of any particular HI structure highlights areas where the different phases dominate in column density. We calculate the column density of HI ($N_{HI}$) with the assumption the HI is optically thin \citep[from][]{roberts1975RadioObservations}:
\begin{equation}
\centering
    N_{HI}(x,y) = 1.823 \times 10^{18}\int T_B(x,y,v) \text{ }dv,
\label{eq:nhgen}
\end{equation}
where $N_{HI}$ is the column density, $T_B$ is the brightness temperature along the positional axes ($x$ and $y$) and velocity axis ($v$).

We take the optically thin regime as a reasonable assumption since these are low column density areas. In the \citet{dempsey2022GASKAPHIPilot} absorption study, they showed that the HI column density correction factor for optical depth is less than 1.05 for uncorrected column densities below $10^{21} \text{ cm}^{-2}$.

As we model $T_B(x,y,\nu)$ as a sum of N Gaussians along the $\nu$ axis, we can simplify Equation \ref{eq:nhgen} that for any component $n$, the column density is:
\begin{equation}
\centering
    N_{n,\text{HI}}(x,y) = 1.823 \times 10^{18}\sqrt{2\pi} A_n(x,y) \sigma_n(x,y),
\label{eq:nhgauss}
\end{equation}
where $A_n$ and $\sigma_n$ are the amplitude and and velocity dispersion respectively of the n$^{\text{th}}$ component. 

Additionally we define the CNM fraction in this work as such:
\begin{equation}
    f_{\text{CNM}} =\frac{N_{\text{(HI,CNM)}}}{N_{\text{(HI,CNM)}}+N_{\text{(HI,UNM)}}+N_{\text{(HI,WNM)}}}.
\label{eq:cnmfrac}
\end{equation}

In Figure \ref{fig:phasedist} in the Alpha cloud, we see that the fraction of CNM decreases from south-west to north-east, with the exception of the cold clump at the north-east end of the cloud (RA 01:09:31, DEC -71:16:47). At these velocities (100-120 km s$^{-1}$) the main emission from the SMC lies to the south-west of this field, so the closer to the the main body of the SMC, the higher the CNM fraction. Additionally, at this position of the Alpha cloud in projection, the emission of the main body of the galaxy is seen at $v>135$ km s$^{-1}$. Looking at the velocity maps in Figure \ref{fig:bigalphafig} the south-west region deviates less from the main body than the north-east. In summation, the further we go from the main body emission, in velocity or spatially, the lower the CNM fraction typically gets. The only area that bucks this trend is the compact cold clump in the north-east. It is centred at a highly deviant velocity from the SMC and has a CNM fraction of $\sim$ 0.6.

The Hook cloud has a stronger contribution from the UNM phase than either of the other two clouds. The maximum value for the CNM fraction is $f_{\text{CNM,max}}=0.67$ and changes as we move around the cloud. The western side of the cloud is where the CNM fraction is highest, especially towards the centre of the filament. This means the CNM is shrouded by an envelope of UNM. The main body emission peaks at 150 km s$^{-1}$ in this field and at these velocities the main body emission lies off to the south and south-west of this field. There is no strong relation between the velocity structure and the CNM fraction, as the CNM fraction seems to decrease in the northern most part, whereas, the velocity structure follows a south to north trend, detailed further in Section \ref{subsec:velgrad}. The CNM fraction seems to increase with the thickness of the filament along the western edge.

The Gamma cloud has strong contributions from the CNM and WNM, with the majority of the CNM residing at the south end of the cloud and the WNM residing at the north end. There is a definite gradient in the CNM fraction as we move up the cloud. It transitions from completely CNM to completely WNM from bottom to top as shown in Figure \ref{fig:phasedist}. In the southern part of this cloud, the CNM fraction approaches 1, which is the highest we record for any of the three clouds. Whereas the other clouds point to a scenario where the CNM is shrouded in a more diffuse WNM envelope, this cloud has no WNM envelope at its southern end. At this clouds velocity (100-112 km s$^{-1}$) the main body emission is south of the cloud and again at the clouds location, the main emission of the SMC starts at 125 km s$^{-1}$ and peaks at 145 km s$^{-1}$. So, similarly to the general trend seen in Alpha cloud, the Gamma cloud has a higher CNM fraction in the area that is closer to the main body of the galaxy. Additionally, as the CNM fraction decreases, the velocity increasingly deviates from the peak emission in this area.

\subsection{HI mass and density}
\label{subsec:himass}

In Table \ref{tab:cloudprops} we show the calculated masses for each cloud separated by phase and in total as well as the CNM volume density.
\begin{table*}
\caption{Physical properties of each cloud, derived as explained in Section \ref{subsec:himass}.}  
\label{tab:cloudprops}      
\centering          
\begin{tabular}{c c c c c c c c c c}
\hline\hline        
\multirow{2}{*}{Cloud} & M$_{\text{CNM}}$ & e$_{\text{M,CNM}}$ & M$_{\text{WNM}}$ & e$_{\text{M,WNM}}$ & M$_{\text{tot}}$ & e$_{\text{M,tot}} $ & $\tilde{W} $ & n$_{\text{CNM}}$ & e$_{\text{n,CNM}}$\\ 
&$(10^4 \text{M}_\odot)$ & $(10^4 \text{M}_\odot)$ & $(10^4 \text{M}_\odot)$ & $(10^4 \text{M}_\odot)$ & $(10^4 \text{M}_\odot)$ & $(10^4 \text{M}_\odot)$ & (pc) & ($\text{cm}^{-3}$) & ($\text{cm}^{-3}$) \\ 
\hline                    
   Alpha & $1.20$ & $0.19$ & $0.21$ & $0.03$ & $1.40$ & $0.19$ & $23.1$ & $5.14$ & $1.38$\\  
   Hook & $1.34$ & $0.21$ & $3.86$ & $0.61$ & $5.21$ & $0.65$ & $33.7$ & $1.76$ & $0.48$\\
   Gamma & $1.33$ & $0.21$ & $0.68$ & $0.11$ & $2.01$ & $0.24$ & $20.6$ & $1.28$ & $0.47$\\
\hline                  
\end{tabular}
\tablefoot{ Columns 1-6 give the CNM, WNM, and total masses of each cloud and their respective uncertainties. Column 7 gives the measured width of the CNM component. Columns 8-9 give the number density of the CNM from Equation \ref{eq:numdensity} and its uncertainty.}
\end{table*} For these structures, calculating an HI volume density is non-trivial. Hyperspectral data gives us no information about the line of sight structure, thus we have no way of empirically measuring how extended these structures are along the line of sight. In previous studies, the HI density has been calculated by assuming that the depth of a structure along the line of sight is the same as the width \citep{for2016DistanceProperties,mcclure-griffiths2006MagneticallyDominated}. We adopt a similar approach. Since these structures are not spherical, we cannot fit a 2D Gaussian to the column density images of these clouds to reliably estimate the width, so we treat them as filamentary structures. To measure the width of the filamentary structures we used the radfil package from \citet{zucker2018RadFilPython}. This package defines a path through a filament and fits a Gaussian or Plummer profile to the mean profile of the filament. To define these paths through the clouds we masked pixels outside the 67\% contour level and passed the masked data through the radfil profile builder. The resulting mean profile is then fit with a Gaussian profile which provides a FWHM that we take as the defining width of the filament. To get a representative density, we then divide the maximum column density at a given position along the filament by the width of the filament in angular units and accounting for the distance of the SMC, as in Equation \ref{eq:numdensity} \citep[as is done in][]{for2016DistanceProperties,benbekhti2006PhysicalProperties}): 

\begin{equation}
\centering
    \text{n}(x) = \frac{N_{HI, max}(x)}{d\tan\theta(x)},
\label{eq:numdensity}
\end{equation}
where $x$ is the distance along the filament, $\theta(x)$ is the FWHM at $x$, and $d$ is the distance to the cloud ($63 \pm 5$ kpc). We used the reported distance of the SMC from \citet{diteodoro2019DynamicsSmall} in this instance, even though it may not be the distance to these individual clouds. We are unable to determine the distance to the clouds themselves without stellar associations that provide this information. The uncertainty inherent in this assumption is translated into the uncertainty in the number density measurement by including the uncertainty in the SMC distance reported in \citet{diteodoro2019DynamicsSmall}. These values and associated uncertainties for the measured filament widths and subsequent number densities are reported in Table \ref{tab:cloudprops}. 

Using this method, we find the minimum values for the FWHM in each filament of around ~67" (20.6 pc) which is just over twice our beamsize of 30". So the thinnest parts of the filaments are barely resolved. This could indicate that the CNM structures are more compact than we are able to measure, thus our measures of density, which range from 1-5 $\text{cm}^{-3}$, should be treated as lower limits. These values are on the lower end of the typical densities expected for the CNM \citep{wolfire1995NeutralAtomic}. For comparison, in one study of CNM filaments in the Milky Way \citet{kalberla2016ColdMilky} calculate a upper limit on filament thickness of 0.3  pc. In another study of HI filaments in the Riegel-Crutcher Cloud \citep{mcclure-griffiths2006MagneticallyDominated} the magnetically dominated filaments had a typical width of 0.1 pc. Using these widths would increase the density to 70-200 $\text{cm}^{-3}$, within the high end of the expected CNM densities. To resolve structures on these scales for the SMC we would need $\leq 1$" angular resolution, which is not readily achievable with current radio interferometers.

\subsection{Velocity structure}
\label{subsec:velgrad}
From the best solutions for each field we can construct velocity maps for each phase. These velocity maps are shown in the middle columns of Figures \ref{fig:bigalphafig}, \ref{fig:bighookfig}, and \ref{fig:biggammafig}.

In the Alpha cloud we find different velocity structures in each component. The CNM component shows a $\sim$ -10 km s$^{-1}$ gradient from west to east whereas the WNM component has no strong gradient from west to east. Both components are centred around the same velocity on the western side, but moving across to the eastern side end up diverging from each other by 12 km s$^{-1}$. This divergence is driven mainly by the CNM velocity gradient. The CNM velocity decreases as we move the side of the cloud furthest from the SMC main body at these velocities. 

The Hook cloud has an interesting velocity structure, it has a gradient that runs south to north with decreasing velocity along both sides. The CNM and UNM follow each other very well, only offset by 0.7 km s$^{-1}$ on average where the CNM lies, along the western edge. The velocity gradient shows that the areas in which the cloud is closest to the main body emission (in the south-west direction) have the least divergent velocities from the main body. This is a trend we see in all 3 clouds, a velocity decrease as we move radially away from the SMC centre in opposition to the general velocity trend we see across the SMC. The SMC emission moves east-west across the sky as the velocity decreases from $\sim$ 200 - 100 km s$^{-1}$.

We see a different trend in the Gamma cloud. Figure \ref{fig:biggammafig} shows a clear velocity gradient from 118 km s$^{-1}$ at the bottom of the cloud to 95 km s$^{-1}$ at the top of the cloud. The CNM and WNM components follow each other along this gradient, south to north along the cloud at a slight offset. They are offset from each other on average by 2.7 km s$^{-1}$. So, they both slow down relative to the SMC main body as the distance from the SMC increases. As mentioned in Section \ref{subsec:phasedist}, the CNM fraction decreases along this same path, so overall in this cloud the warmer the HI, the slower the LOS velocity relative to the main body emission. 

\subsubsection{Deviation velocities}

To make a measure of the deviation velocity of each cloud, a measure commonly used to categorise low, intermediate, and high velocity clouds (LVC, IVC, and HVCs) in the Milky Way, we calculate the first moment map of the galaxy with the clouds removed. To remove the clouds, we mask every voxel within the fields analysed where the total emission from the ROHSA model is above the noise level. From this masked PPV cube, we obtain the first moment and compare the mean velocities of each cloud with the first moment velocity at their position. We define the deviation velocity ($v_{\text{DEV}}$) as in Equation \ref{eq:devvel} below.

\begin{equation}
\label{eq:devvel}
    v_{\text{DEV}} = v_{\text{peak}} - v_{\text{M1}}.
\end{equation}
where $v_{\text{peak}}$ is the peak velocity, the velocity at which the brightness temperature from the ROHSA model is at its maximum,
and $v_{\text{M1}}$ is the first moment velocity. From Equation \ref{eq:devvel} we find that the deviation velocities of the Alpha, Hook, and Gamma clouds are -56.4, -13.8, and -34.8 km s$^{-1}$ respectively.

By the definitions asserted in \citet{wakker2004HVCIVC}, classifying clouds by their deviation velocity, the Alpha cloud is an IVC and the Hook and Gamma clouds are LVCs of the SMC.

\subsubsection{Clouds in the context of the 3D morphology of the SMC}
\label{subsec:Murray2024}
Recently, work was done in \citet{murray2024GalacticEclipse} to investigate the distribution of the SMC HI gas along the line of sight, using distances obtained from the stellar population. They found that the stellar population of the SMC could be separated into two distinct groups, sitting at two distinct distances along the line of sight and used this information to separate the HI into 'front' and 'behind' sections. The distinction between these two components is well-determined where there are many stellar objects, which is primarily in the main body of the galaxy. The distinction becomes hazier in the outskirts of the galaxy, where there are fewer stellar objects. We compared the central velocities of the clouds in this work to the first moment maps of the front and behind components constructed in \citet{murray2024GalacticEclipse}. 

From the first moment maps in \citet{murray2024GalacticEclipse} we extracted the regions around each cloud in this work. From each subsection we calculated the median value in the front and behind components, obtaining two respective first moment velocities for each cloud region. To characterise the uncertainty on these first moment velocities, we measured the dispersion in each field. The range of first moment velocities in these fields were quite extreme, due to low column density lines of sight, so the reported uncertainty is the mean of the dispersion measured when the 10\% most positive and most negative values are excluded, and the dispersion measured when lines of sight with column densities below $1.32 \times 10^{20} \text{ cm}^{-2}$ are excluded (thus above the 3$\sigma$ column density sensitivity limit reported in \citet{pingel2022GASKAPHIPilot}). 

For the Alpha field, the median velocities for the front and behind components are $166.1 \pm 5.2$ km s$^{-1}$ and $120.8 \pm 22.4$ km s$^{-1}$ respectively. For the Hook field, the median velocities for the front and behind components are $142.1 \pm 22.8$ km s$^{-1}$ and $143.3 \pm 14.7$ km s$^{-1}$ respectively. For the Gamma field, the median velocities for the front and behind components are $142.6 \pm 15.6$ km s$^{-1}$ and $146.1 \pm 11.0$ km s$^{-1}$ respectively. The difference between the first moment front and behind velocities in the Hook and Gamma cloud fields is relatively small, eclipsed by the uncertainties on each measurement. There is a larger distinction between the two component velocities for the Alpha cloud field, however the difference between them is still within a $2\sigma$ uncertainty level. Given that the peak velocity of the Alpha cloud is 106.4 km s$^{-1}$, it is offset from front component by 59.7 km s$^{-1}$, a 12$\sigma$ level, but offset by 14.4 km s$^{-1}$ and well within the 1$\sigma$ level of the behind component. This suggests that the Alpha cloud belongs to the behind component of the SMC identified in \citet{murray2024GalacticEclipse}. 

Working under that assumption may shed some light on the distance to the Alpha cloud. In Section \ref{subsec:himass} we assumed a distance to all 3 clouds of $63 \pm 5$ kpc from \citet{diteodoro2019DynamicsSmall}, but if the Alpha cloud belongs to the behind component of the SMC, we could also assume it sits at the mean distance of that component reported in \citet{murray2024GalacticEclipse} of $66$ kpc. If this is true, the physical scales of the Alpha cloud will increase by 5\% and the HI mass will increase by 10\%.

\section{Discussion}
\label{sec:discussion}

\subsection {CO comparison}
\label{subsec:cocomp}
$^{12}$CO$(2\rightarrow1)$ data were obtained during targeted APEX observations by \citet{diteodoro2019MolecularGas}. It covered the entirety of the Alpha cloud and a subsection of the Hook cloud field.

\begin{figure*}[h]
  \centering
  \begin{subfigure}{0.9\textwidth}
    \includegraphics[width=\textwidth]{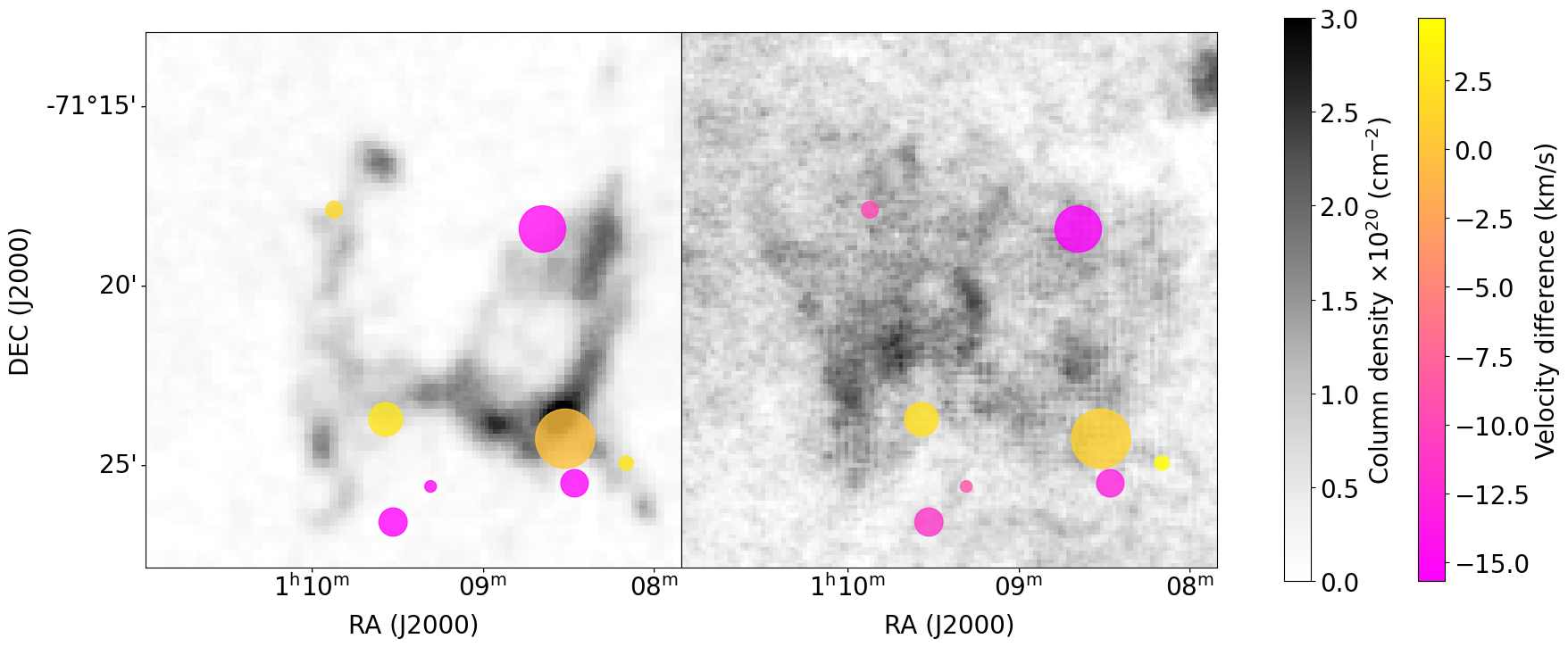}
  \end{subfigure}
  \begin{subfigure}{0.9\textwidth}
    \includegraphics[width=\textwidth]{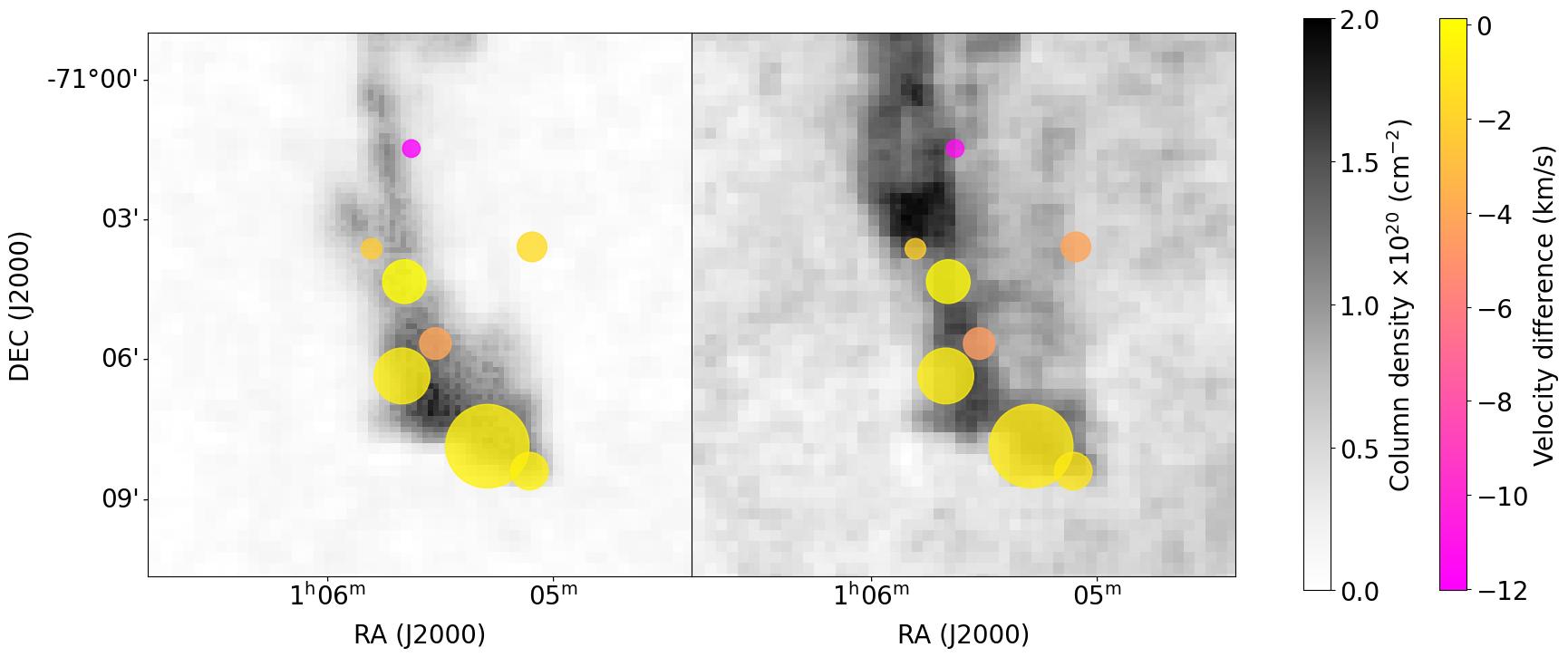}
  \end{subfigure}
  \caption{CO velocity comparison ($v_{\text{CO}}-v_{\text{HI}}$) for Alpha cloud (Top) and Hook cloud (Bottom). Size of circle markers indicate the strength of the integrated CO flux density scaled linearly.}
  \label{fig:COcomp}
\end{figure*}

Within the field encompassing the Alpha cloud, there were 8 clumps of CO detected in \citet{diteodoro2019MolecularGas}. Comparison of the HI CNM component central velocities and the CO data show that four of the clumps agree in velocity within a 3 channel ($\sim$3 km s$^{-1}$) window with the CNM, whereas the other four are all offset by approximately 15 km s$^{-1}$ from the CNM, see Figure \ref{fig:COcomp}. The four clumps that are consistent with with the velocity structure of the CNM, are more spatially coincident with the high column density regions of CNM than the others. This divides these clumps into two distinct populations, one that likely belongs to the Alpha cloud and one that does not. Interestingly, the clumps of CO that lie off the highest column density areas of this feature we consider here can not be matched with any strong HI emission. There is very little HI emission in this region at $v< 93 \text{ km s}^{-1}$ that is significant (at 3$\sigma$ above the noise). 

When comparing the CO data to the HI WNM component central velocities, they agree within the same velocity window as the CNM component, except on the eastern side. This is due to the divergence of the CNM from the WNM in velocity space, which we note in Section \ref{subsec:velgrad}. From this, we conclude that the CO is more dynamically aligned with the CNM than the WNM.

Within the field encompassing the Hook cloud there were nine clumps detected in CO. This cloud spans three times the size of the Alpha cloud in both spatial dimensions, so the entirety of the Hook cloud was not observed in CO APEX observations from \citet{diteodoro2019MolecularGas}. The observations cover the areas of strong total HI emission in the base of the western edge of the cloud, shown in Figure \ref{fig:COcomp}. Seven out of nine clumps are in agreement with the CNM within 3 km s$^{-1}$. These same seven clumps are offset slightly more from the UNM velocities. The two clumps that strongly disagree with this trend are slightly spatially offset to the CNM filament, however one clump that agrees with the CNM velocities is also offset from the CNM filament. These two anomalies are also among the clumps with the lowest integrated flux densities. Clumps with higher integrated flux densities and more spatially coincident with the high CNM column density regions, agree best with the central velocities of the CNM component. This indicates that the strongest CO clumps are associated with the dense CNM regions of the Hook cloud. Since there is little difference in the velocities of the UNM and CNM, we could just as equally suppose that the CO is associated with the UNM envelope so that all this gas is travelling together.

Overall, in both clouds, the CO is more dynamically aligned with the CNM, significantly in the Alpha cloud and marginally in the Hook cloud, particularly in areas with high CNM column density. 

\subsection{Phase velocities}
\begin{figure*}[h]
\centering
\begin{subfigure}{0.95\textwidth}
  \includegraphics[width=\textwidth]{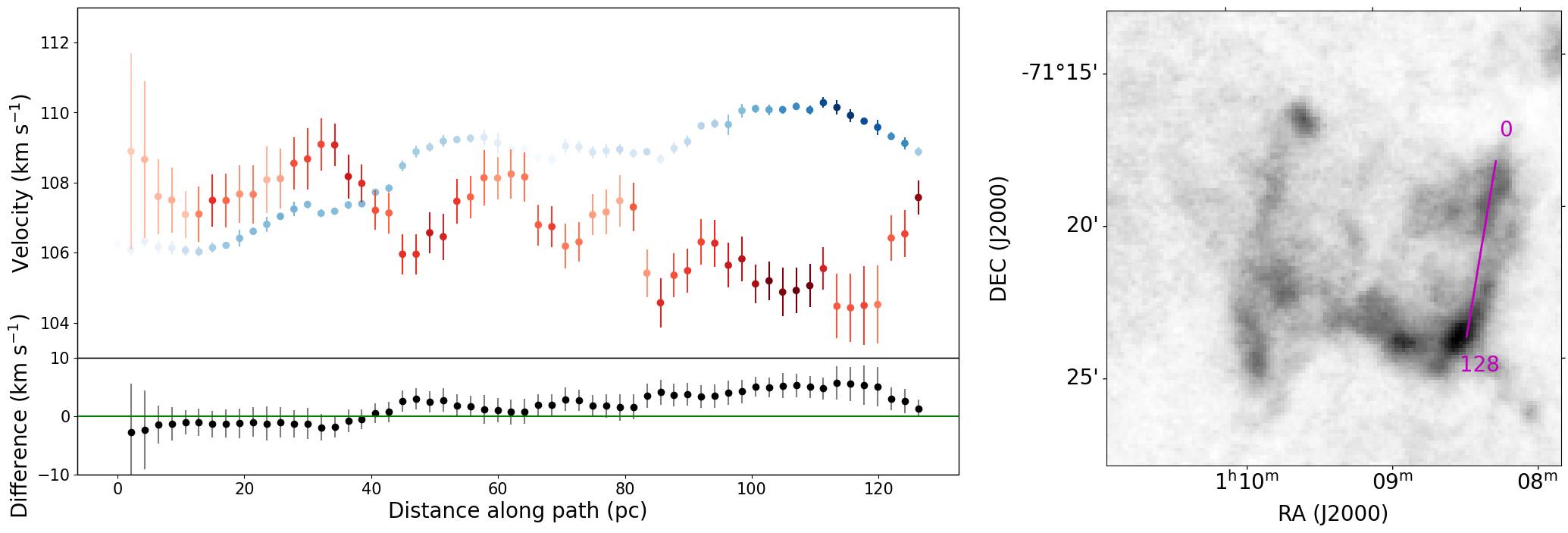}
\end{subfigure}
\begin{subfigure}{0.95\textwidth}
  \includegraphics[width=\textwidth]{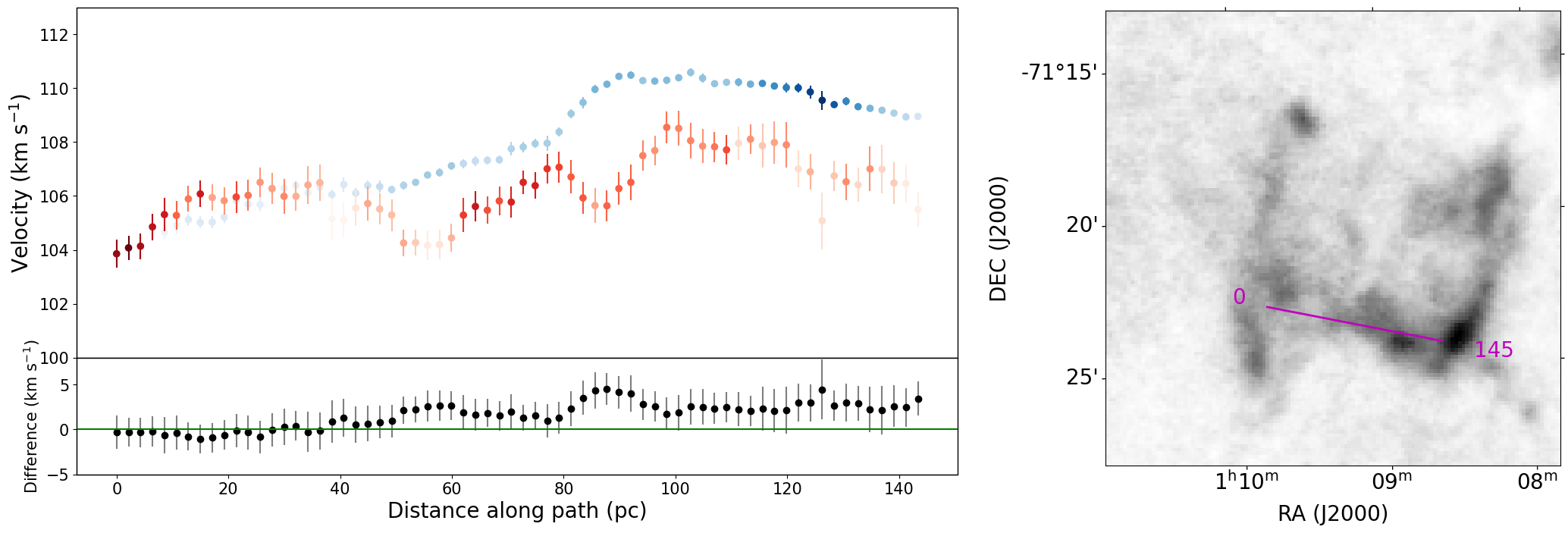}
\end{subfigure}
\caption{Left: Fitted component central velocities for the different phases for two different paths through the Alpha cloud (top and bottom). The WNM is shown in red, the CNM in blue with the darker colours indicating higher column density. The error bars in the top panels show the 1$\sigma$ uncertainties from Section \ref{subsec:errorestimate}. The error bars in the bottom panels show the combined uncertainties at the 3$\sigma$ level. Right: Path through the cloud for respective velocity relations shown in purple. 0 denotes the start of the path, and 128 and 145 denote the ends of each path.}
\label{fig:alphaphasevel}
\end{figure*}

With the information obtained from the ROHSA decomposition, it is possible to trace the velocity of each phase through the cloud. We outlined the trends in the velocity gradients of the individual phases in Section \ref{subsec:velgrad}, but discuss here the relationship between the two different velocity gradients in different parts of each cloud. In the Alpha cloud when plotting different paths through the cloud towards the south-west end, an offset between the two phases becomes apparent. In Figure \ref{fig:alphaphasevel} we show two paths through the Alpha cloud, one from east to west and another from north to south. Both paths terminate at the concentration of emission in the south-west of the cloud and show a significant offset in the velocities of the two phases. This offset is most obvious in Figure \ref{fig:alphaphasevel} (b) where it reaches $\sim$ 5 km s$^{-1}$. This offset is also seen in a region of high column density for both phases with relatively low uncertainties on the central velocity, providing strong evidence for a velocity offset between the phases towards this region of the cloud.

For the Hook cloud, there is very little difference in the central velocities of each phase, and as such, there is no significant consistent or increasing offset between the two phases as we move along the curved filament that defines the Hook cloud. There is a slight offset of less than 1 km s$^{-1}$ in the densest area of the cloud towards the southern end, but this offset is within the 3$\sigma$ uncertainty window. 

Finally, for the Gamma cloud the velocity difference between the phases is a little more complex. There are two CNM components that need to be averaged to obtain a representative central velocity for the total CNM. Both of these CNM components have higher uncertainties than the CNM components of the other two clouds. This is likely because these components have similar central velocities and were easily confused with each other amongst the many ROHSA runs completed for the bootstrapping uncertainty method described in Section \ref{subsec:errorestimate}. These larger uncertainties are compounded when taking the mean of the two values and this makes it difficult to have confidence in offsets of the same magnitude as in the Alpha cloud. Additionally, the measure of the offset becomes less meaningful when the column density of either phase falls below the noise limit which poses problems for this cloud. The CNM is most prominent at the southern end of the cloud and the WNM at the northern end, so there is not much area where the phases overlap at high column densities. There are a few points in the northern end of the cloud which have significant CNM column densities where an offset of $\sim$4 km s$^{-1}$ can be measured at levels exceeding 3$\sigma$. From just these points, it cannot be concluded that there is a consistent offset between the phases, but it suggests that an offset is present at least in the northern end of the cloud.

\subsection{Factors affecting morphology}

\subsubsection{Alpha cloud}
The Alpha cloud has a complex structure, the most irregular of the three clouds considered in this work. It does not have just one concentration of cold material, with the north-east and south-west ends of the cloud each containing significant amounts of CNM. This cloud was also analysed in \citet{pingel2022GASKAPHIPilot}, where they looked at the mean properties of the cloud. When they took a mean spectrum of the cloud and fit a Gaussian they obtained a $\sigma$ value of 2.93 $\text{km s}^{-1}$, which is higher than our obtained value of 1.97 $\text{km s}^{-1}$. This discrepancy is likely due to the presence of the WNM uncovered in this work which would broaden the mean profile. Additionally, in this work we found that the CNM and WNM are not consistently offset from each other in velocity, so this would also add to the broadening of the mean spectrum. These both explain the broader dispersion derived in \citet{pingel2022GASKAPHIPilot} compared to the CNM dispersion in this work.

The Alpha cloud also has an obvious cavity on the western side of the field. This was catalogued as an HI shell in \citet{staveley-smith1997HIAperture} with a heliocentric velocity of 117.7 km s$^{-1}$, corresponding to a local standard of rest (LSR) velocity of 108.7 km s$^{-1}$. We measure a maximum velocity of 113.4 km s$^{-1}$ (LSR) for the CNM component in this field which is consistent with the shell velocity considering our velocity resolution of 0.98 km s$^{-1}$ and the velocity resolution in \citet{staveley-smith1997HIAperture} of 1.6 km s$^{-1}$. This cavity was also noted in the analysis in \citet{pingel2022GASKAPHIPilot} where there was no conclusive stellar association made due to the absence of radial velocity information from the Gaia release \citep{gaiacollaboration2021GaiaEarly}. An expanding shell could explain the absence of HI, specifically the CNM, in this region, but it does not explain the offset in velocity between the phases seen in the south-western side of the shell, where the column density is highest. The offset in velocity, coupled with the higher CNM fraction at this end of the cloud, suggests stellar feedback from the direction the main galaxy is acting on the cloud and has managed to move the less dense WNM more efficiently than the CNM. This feedback has begun to strip the cloud of its WNM envelope at this position, leaving the CNM and CO behind, similar to the head-tail structure seen in some HVCs and IVCs. 

\begin{figure}[h]
    \centering
    \includegraphics[width=0.5\textwidth]{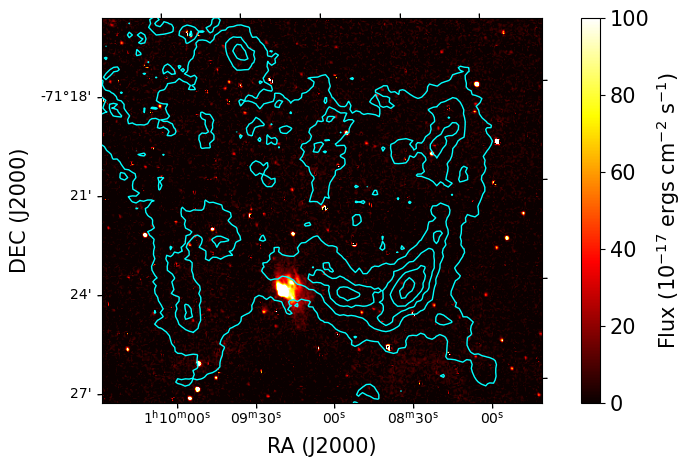}
    \caption{MCELS H$\alpha$ data with the HI column density of the Alpha cloud overlaid in contour levels from $1\times 10^{20}$ to $4.2\times 10^{20} \text{ cm}^{-2}$.}
    \label{fig:alphaHalpha}
\end{figure}

An additional piece of information available in this area is the H$\alpha$ data from the Magellanic Cloud Emission Line Survey (MCELS) \citep{winkler2015InterstellarMedium}. Figure \ref{fig:alphaHalpha} shows the MCELS H$\alpha$ where there is strong emission in the southern region of the cloud. Around this H$\alpha$ emission, there is an enhancement of the CNM fraction, except on the southern side where there is very little HI. The positional coincidence of this H$\alpha$ emission with the HI cloud suggests they could be related, but there is no spectroscopic data available for this region at sufficient resolution to make an association between the cloud and the H$\alpha$ emission.  

Overall, the Alpha cloud seems to be shaped by multiple processes. Star formation that is facilitating expansion, or turbulent motions as suggested in \citet{pingel2022GASKAPHIPilot}, have created a cavity in the centre of the cloud, while stellar winds from the main galaxy have stripped the WNM envelope in the south-western head of the cloud. Additionally, compact H$\alpha$ emission in the south of the cloud could indicate stellar activity may be ionising the surrounding HI, if it is associated with the Alpha cloud.

\subsubsection{Hook cloud}
The Hook cloud seems to have a simpler morphology than the Alpha cloud, with the CNM and UNM matching each other well in position and velocity. There is no velocity gradient along the length of the cloud as we move further away from the main body of the galaxy. This suggests that the filament is not strongly affected by star formation from within the SMC main body. As it has a concentration of CNM on the western side, with a UNM envelop that also loops around to the eastern side, it seems more likely to be caused by an expansive force from the centre of the field. It is not spherical, which could be due to expansion occurring at multiple points within this area, causing uneven elongation along the different axes.

The source of this expansive force could come from stellar objects in the area. \citet{martinez-delgado2019NatureShell} identified an arm of stellar objects that extends into this field that could be injecting energy into the diffuse medium of the SMC outskirts, creating CNM filaments. There are multiple stellar clusters within this field, catalogued in \citet{bica2020UpdatedSmall}, which could provide the energy needed to create a shell. However, all of these clusters, for which we have age measurements, are old ($t>100$ Myr) and no shell in the most recent catalogue of HI shells in the SMC \citep{staveley-smith1997HIAperture} exceeds 40 Myr in age. We can characterise the energies associated with the potential shell by using the information from the HI. We are unable to measure an expansion velocity of this shell-like structure, as there are no apparent front and back walls in HI spectra from the centre of the field. However, this does not preclude the existence of front and back walls as they may be below the noise limit. So for the Hook cloud, we therefore assume expansion velocities of $v_{\text{exp}}=10-20$ km s$^{-1}$, as this is within the range found in \citet{staveley-smith1997HIAperture} for the larger shells. Using the total HI mass shown in Table \ref{tab:cloudprops} of $5.21 \text{ M}_{\odot}$, we calculate the kinetic energy of the shell to be $0.5-2.1 \times 10^{50}$ erg. 

We can also define a kinetic age ($t_k$), where $t_k = R_s/v_{\text{exp}}$, using the mean of the two shell axes as the representative radius of the shell ($R_s$). The two axes of the shell measure at 360 and 680 pc, thus $R_s=520$ pc, giving the kinetic age $t_k=25-51$ Myr. Following the formalism in \citet{mccray1987SupershellsPropagating} we can also calculate a timescale and number of stars required to produce this shell if there was constant supernovae energy injection using $R_s = 97\text{ pc} \left(N_\ast/n_0\right)^{1/5}t_7^{3/5}$ and $v_{\text{exp}} = 5.7$ km s$^{-1}\left(N_\ast/n_0\right)^{1/5}t_7^{-2/5}$, where $t_7$ is the time in units of 10 Myr, $N_\ast$ is the number of massive stars, and $n_0$ is the ambient density of the medium before the shell was formed. To estimate the ambient density, we take the total mass of the HI and divide by the volume of shell, using the minor axis of the shell as the depth along the line of sight. This gives an ambient density of 0.05 cm$^{-3}$. With $R_s=520$ pc, $n_0 = 0.05 $ cm$^{-3}$, and $v_{\text{exp}}=10-20$ km s$^{-1}$, we get $t_7=1.5-3.1$ and $N_\ast=7-57$. This range in timescale overlaps with the range we derived for the dynamic timescale. 

These measurements are compatible with the Hook cloud being a shell formed within the last 50 Myr from a cluster of massive stars. The elongated expansion along the major axis of the shell is reminiscent of HI chimney structures, such as in \citet{normandeau1996GalacticChimney}, \citet{pidopryhora2007OphiuchusSuperbubble}, and \citet{dawson2008CarinaFlare}. In these cases the HI shell expanded much further away from the Galactic Plane of the Milky Way than they could into the Galactic Plane due to the density and pressure difference. It is also common to see a cap of HI that forms the top of the chimney, which could be the northern edge of the Hook in this case.

The MCELS data for this field shows low-level H$\alpha$ emission along the southern edges of the Hook on both sides which was also highlighted in \citet{mcclure-griffiths2018ColdGas}, but as this field is at the edge of the MCELS data range, we have no information about the H$\alpha$ emission for the cap of the Hook.
An HI chimney-like expansion history seems the most probable explanation for the Hook cloud, which would make it one of the largest shells identified in the SMC, when comparing with the catalogue presented in \citet{staveley-smith1997HIAperture} and expanded upon in \citet{hatzidimitriou2005PropertiesHI}.

\subsubsection{Gamma cloud}
The Gamma cloud exhibits a head-tail structure that is commonly seen in HVCs. This could indicate that the cloud is experiencing infall, with simulations in \citet{heitsch2022MassMorphing}, \citet{konz2002DynamicalEvolution} and \citet{quilis2001WhereAre} all showing that the heads of infalling clouds are typically colder than the trailing tail of the cloud. This is consistent with what we see in the Gamma cloud and is further supported by the fact that the cloud is elongated roughly along the axis that points towards the dynamical centre of the SMC, indicating the path it has taken. However, this would also be seen if a neutral cloud that already existed in the periphery was subject to ram-pressure stripping from wind from the main galaxy. The cloud would be elongated along the direction of the stellar feedback, and the cold core would remain at the base of the cloud while the WNM is stripped and pushed away to trail behind the cold core. In the outflow scenario, we are catching the cloud at a point before the unshielded CNM core has begun to dissociate. This is similar to the outflow scenario proposed in \citet{noon2023DirectObservations} for smaller Milky Way clouds. The clouds analysed in \citet{noon2023DirectObservations} also had varying amounts of CO present, depending on the age of the cloud, which was not available for the Gamma cloud in this analysis.

\begin{figure}[h]
    \centering
    \includegraphics[width=0.35\textwidth]{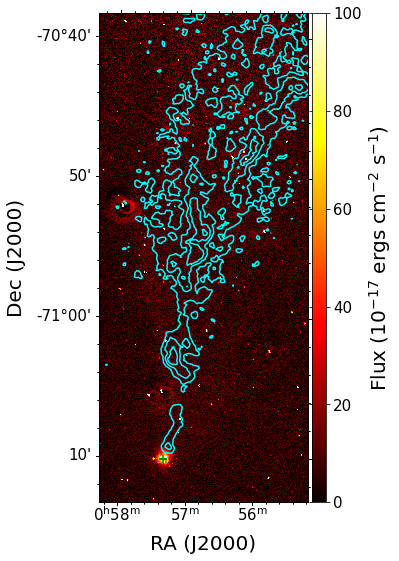}
    \caption{MCELS H$\alpha$ data with the HI column density of the Gamma cloud contour levels from $7\times 10^{19}$ to $42\times 10^{19} \text{ cm}^{-2}$. Green marker shows the position of HW 32 a young stellar cluster of the SMC.}
    \label{fig:gammaHalpha}
\end{figure}

As with the Alpha cloud, the MCELS data shows that there is significant H$\alpha$ emission at the base of the Gamma cloud, shown in Figure \ref{fig:gammaHalpha}, that is also spatially conicident with a stellar cluster HW32, catalogued in \citet{maia2014MassDistribution}. There is no radial velocity information available at sufficient resolution to confirm a dynamic association with the Gamma cloud, but the spatial coincidence is compelling. This region could be the source of the stellar activity that is ionising and pushing the WNM away from the base of HI cloud. However, the cluster would need to provide sufficient energy to elongate an HI cloud over 560pc. So a combination of feedback from this cluster and the SMC could be a possible formation scenario.

\section{Conclusions}
\label{sec:conclusions}
From this work, we have found that in the periphery of the SMC, there is a significant amount of cold material in cloud structures detached from the main body of the SMC. This result complements the results of previous absorption surveys of the SMC in building a picture of the CNM distribution for the SMC. Our main results are:
\begin{itemize}
    \item significant maximum values of the CNM fraction ($f_{\text{CNM}}$), ranging from 0.6-1 across the three clouds in a low metallicity environment;
    \item preferential spatial and dynamical agreement of observed CO clumps with the CNM over the WNM;
    \item new mass measurements for each cloud showing they all have similar masses, of the order of $10^4 \text{M}_{\odot}$, in line with the results from \citet{mcclure-griffiths2018ColdGas}; and
    \item first estimates of the CNM number density providing a lower limit of the order 1 cm$^{-3}$ for these clouds, limited by spatial resolution.
\end{itemize}
Our analysis has also shown that the Alpha and Gamma clouds are likely shaped by stellar feedback from the direction of the main body of the SMC, as well as possible nearby stellar clusters, stripping the WNM envelope away from the shielded CNM at the bases of these clouds. However, we cannot exclude the infall scenario for the Gamma cloud. In contrast to the other two clouds, the Hook cloud is likely an HI supershell showing signs of chimney-like blowout, one of the largest shells catalogued in the SMC.

\begin{acknowledgements}
We would like to thank the anonymous referee for their comments that helped improve the quality of this work.

We would like to acknowledge the Interstellar Institute programs II5 and II6 for fostering collaboration and discussion around the work presented in this paper.

This scientific work uses data obtained from Inyarrimanha Ilgari Bundara / the Murchison Radio-astronomy Observatory. We acknowledge the Wajarri Yamaji People as the Traditional Owners and native title holders of the Observatory site. CSIRO’s ASKAP radio telescope is part of the Australia Telescope National Facility (https://ror.org/05qajvd42). Operation of ASKAP is funded by the Australian Government with support from the National Collaborative Research Infrastructure Strategy. ASKAP uses the resources of the Pawsey Supercomputing Research Centre. Establishment of ASKAP, Inyarrimanha Ilgari Bundara, the CSIRO Murchison Radio-astronomy Observatory and the Pawsey Supercomputing Research Centre are initiatives of the Australian Government, with support from the Government of Western Australia and the Science and Industry Endowment Fund.

This paper includes archived data obtained through the CSIRO ASKAP Science Data Archive, CASDA (http://data.csiro.au).

EMDT was supported by the European Research Council (ERC) under grant agreement no. 101040751.

This work made use of the numpy \citet{harris2020ArrayProgramming} and matplotlib \citet{hunter2007Matplotlib2D} python packages. Additionally, this work made use of Astropy:\url{http://www.astropy.org} a community-developed core Python package and an ecosystem of tools and resources for astronomy \citep{astropycollaboration2013AstropyCommunity, astropycollaboration2018AstropyProject,astropycollaboration2022AstropyProject}.

\end{acknowledgements}

\bibliographystyle{aa}
\bibliography{refs.bib}

\begin{thebibliography}{54}
\expandafter\ifx\csname natexlab\endcsname\relax\def\natexlab#1{#1}\fi

\bibitem[{{Astropy Collaboration} {et~al.}(2022){Astropy Collaboration},
  {Price-Whelan}, Lim, Earl, Starkman, Bradley, Shupe, Patil, Corrales,
  Brasseur, N{\"o}the, Donath, Tollerud, Morris, Ginsburg, Vaher, Weaver,
  Tocknell, Jamieson, {van Kerkwijk}, Robitaille, Merry, Bachetti, G{\"u}nther,
  Aldcroft, {Alvarado-Montes}, Archibald, B{\'o}di, Bapat, Barentsen,
  Baz{\'a}n, Biswas, Boquien, Burke, Cara, Cara, Conroy, Conseil, Craig, Cross,
  Cruz, D'Eugenio, Dencheva, Devillepoix, Dietrich, Eigenbrot, Erben, Ferreira,
  {Foreman-Mackey}, Fox, Freij, Garg, Geda, Glattly, Gondhalekar, Gordon,
  Grant, Greenfield, Groener, Guest, Gurovich, Handberg, Hart,
  {Hatfield-Dodds}, Homeier, Hosseinzadeh, Jenness, Jones, Joseph, Kalmbach,
  Karamehmetoglu, Ka{\l}uszy{\'n}ski, Kelley, Kern, Kerzendorf, Koch, Kulumani,
  Lee, Ly, Ma, MacBride, Maljaars, Muna, Murphy, Norman, O'Steen, Oman,
  Pacifici, Pascual, {Pascual-Granado}, Patil, Perren, Pickering, Rastogi,
  Roulston, Ryan, Rykoff, Sabater, Sakurikar, Salgado, Sanghi, Saunders,
  Savchenko, Schwardt, {Seifert-Eckert}, Shih, Jain, Shukla, Sick, Simpson,
  Singanamalla, Singer, Singhal, Sinha, Sip{\H o}cz, Spitler, Stansby,
  Streicher, {\v S}umak, Swinbank, Taranu, Tewary, Tremblay, {de Val-Borro},
  Van~Kooten, Vasovi{\'c}, Verma, {de Miranda Cardoso}, Williams, Wilson,
  Winkel, {Wood-Vasey}, Xue, Yoachim, Zhang, Zonca, \& {Astropy Project
  Contributors}}]{astropycollaboration2022AstropyProject}
{Astropy Collaboration}, {Price-Whelan}, A.~M., Lim, P.~L., {et~al.} 2022, ApJ,
  935, 167

\bibitem[{{Astropy Collaboration} {et~al.}(2018){Astropy Collaboration},
  {Price-Whelan}, Sip{\H o}cz, G{\"u}nther, Lim, Crawford, Conseil, Shupe,
  Craig, Dencheva, Ginsburg, VanderPlas, Bradley, {P{\'e}rez-Su{\'a}rez}, {de
  Val-Borro}, Aldcroft, Cruz, Robitaille, Tollerud, Ardelean, Babej, Bach,
  Bachetti, Bakanov, Bamford, Barentsen, Barmby, Baumbach, Berry, Biscani,
  Boquien, Bostroem, Bouma, Brammer, Bray, Breytenbach, Buddelmeijer, Burke,
  Calderone, Cano~Rodr{\'i}guez, Cara, Cardoso, Cheedella, Copin, Corrales,
  Crichton, D'Avella, Deil, Depagne, Dietrich, Donath, Droettboom, Earl, Erben,
  Fabbro, Ferreira, Finethy, Fox, Garrison, Gibbons, Goldstein, Gommers, Greco,
  Greenfield, Groener, Grollier, Hagen, Hirst, Homeier, Horton, Hosseinzadeh,
  Hu, Hunkeler, Ivezi{\'c}, Jain, Jenness, Kanarek, Kendrew, Kern, Kerzendorf,
  Khvalko, King, Kirkby, Kulkarni, Kumar, Lee, Lenz, Littlefair, Ma, Macleod,
  Mastropietro, McCully, Montagnac, Morris, Mueller, Mumford, Muna, Murphy,
  Nelson, Nguyen, Ninan, N{\"o}the, Ogaz, Oh, Parejko, Parley, Pascual, Patil,
  Patil, Plunkett, Prochaska, Rastogi, Reddy~Janga, Sabater, Sakurikar,
  Seifert, Sherbert, {Sherwood-Taylor}, Shih, Sick, Silbiger, Singanamalla,
  Singer, Sladen, Sooley, Sornarajah, Streicher, Teuben, Thomas, Tremblay,
  Turner, Terr{\'o}n, {van Kerkwijk}, {de la Vega}, Watkins, Weaver, Whitmore,
  Woillez, Zabalza, \& {Astropy
  Contributors}}]{astropycollaboration2018AstropyProject}
{Astropy Collaboration}, {Price-Whelan}, A.~M., Sip{\H o}cz, B.~M., {et~al.}
  2018, AJ, 156, 123

\bibitem[{{Astropy Collaboration} {et~al.}(2013){Astropy Collaboration},
  Robitaille, Tollerud, Greenfield, Droettboom, Bray, Aldcroft, Davis,
  Ginsburg, {Price-Whelan}, Kerzendorf, Conley, Crighton, Barbary, Muna,
  Ferguson, Grollier, Parikh, Nair, Unther, Deil, Woillez, Conseil, Kramer,
  Turner, Singer, Fox, Weaver, Zabalza, Edwards, Azalee~Bostroem, Burke, Casey,
  Crawford, Dencheva, Ely, Jenness, Labrie, Lim, Pierfederici, Pontzen, Ptak,
  Refsdal, Servillat, \& Streicher}]{astropycollaboration2013AstropyCommunity}
{Astropy Collaboration}, Robitaille, T.~P., Tollerud, E.~J., {et~al.} 2013,
  A\&A, 558, A33

\bibitem[{Ben~Bekhti {et~al.}(2006)Ben~Bekhti, Br{\"u}ns, Kerp, \&
  Westmeier}]{benbekhti2006PhysicalProperties}
Ben~Bekhti, N., Br{\"u}ns, C., Kerp, J., \& Westmeier, T. 2006, A\&A, 457, 917

\bibitem[{Besla {et~al.}(2012)Besla, Kallivayalil, Hernquist, {van der Marel},
  Cox, \& Kere{\v s}}]{besla2012RoleDwarf}
Besla, G., Kallivayalil, N., Hernquist, L., {et~al.} 2012, MNRAS, 421, 2109

\bibitem[{Bica {et~al.}(2020)Bica, Westera, Kerber, Dias, Maia, Santos, Barbuy,
  \& Oliveira}]{bica2020UpdatedSmall}
Bica, E., Westera, P., Kerber, L. d.~O., {et~al.} 2020, AJ, 159, 82

\bibitem[{Boothroyd {et~al.}(2011)Boothroyd, Blagrave, Lockman, Martin,
  Pinheiro~Gon{\c c}alves, \& Srikanth}]{boothroyd2011AccurateGalactic}
Boothroyd, A.~I., Blagrave, K., Lockman, F.~J., {et~al.} 2011, A\&A, 536, A81

\bibitem[{Dawson(2013)}]{dawson2013SupershellMolecular}
Dawson, J.~R. 2013, PASA, 30, e025

\bibitem[{Dawson {et~al.}(2008)Dawson, Mizuno, Onishi, {McClure-Griffiths}, \&
  Fukui}]{dawson2008CarinaFlare}
Dawson, J.~R., Mizuno, N., Onishi, T., {McClure-Griffiths}, N.~M., \& Fukui, Y.
  2008, MNRAS, 387, 31

\bibitem[{Dempsey {et~al.}(2022)Dempsey, {McClure-Griffiths}, Murray, Dickey,
  Pingel, Jameson, D{\'e}nes, {van Loon}, Leahy, Lee, Stanimirovi{\'c}, Breen,
  {Buckland-Willis}, Gibson, Imai, Lynn, \&
  Tremblay}]{dempsey2022GASKAPHIPilot}
Dempsey, J., {McClure-Griffiths}, N.~M., Murray, C., {et~al.} 2022, PASA, 39,
  e034

\bibitem[{Di~Teodoro {et~al.}(2019{\natexlab{a}})Di~Teodoro,
  {McClure-Griffiths}, De~Breuck, Armillotta, Pingel, Jameson, Dickey, Rubio,
  Stanimirovi{\'c}, \& {Staveley-Smith}}]{diteodoro2019MolecularGas}
Di~Teodoro, E.~M., {McClure-Griffiths}, N.~M., De~Breuck, C., {et~al.}
  2019{\natexlab{a}}, ApJ, 885, L32

\bibitem[{Di~Teodoro {et~al.}(2019{\natexlab{b}})Di~Teodoro,
  {McClure-Griffiths}, Jameson, D{\'e}nes, Dickey, Stanimirovi{\'c},
  {Staveley-Smith}, Anderson, Bunton, Chippendale, {Lee-Waddell}, MacLeod, \&
  Voronkov}]{diteodoro2019DynamicsSmall}
Di~Teodoro, E.~M., {McClure-Griffiths}, N.~M., Jameson, K.~E., {et~al.}
  2019{\natexlab{b}}, MNRAS, 483, 392

\bibitem[{Diaz \& Bekki(2011)}]{diaz2011ConstrainingOrbital}
Diaz, J. \& Bekki, K. 2011, MNRAS, 413, 2015

\bibitem[{Dickey {et~al.}(2000)Dickey, Mebold, Stanimirovic, \&
  {Staveley-Smith}}]{dickey2000ColdAtomic}
Dickey, J.~M., Mebold, U., Stanimirovic, S., \& {Staveley-Smith}, L. 2000, ApJ,
  536, 756

\bibitem[{Draine(2011)}]{draine2011PhysicsInterstellar}
Draine, B.~T. 2011, Physics of the {{Interstellar}} and {{Intergalactic
  Medium}} (Princeton University Press)

\bibitem[{For {et~al.}(2016)For, {Staveley-Smith}, {McClure-Griffiths},
  Westmeier, \& Bekki}]{for2016DistanceProperties}
For, B.~Q., {Staveley-Smith}, L., {McClure-Griffiths}, N.~M., Westmeier, T., \&
  Bekki, K. 2016, MNRAS, 461, 892

\bibitem[{{Gaia Collaboration} {et~al.}(2021){Gaia Collaboration}, Luri,
  Chemin, Clementini, Delgado, McMillan, {Romero-G{\'o}mez}, Balbinot,
  {Castro-Ginard}, Mor, Ripepi, Sarro, Cioni, Fabricius, Garofalo, Helmi,
  Muraveva, Brown, Vallenari, Prusti, {de Bruijne}, Babusiaux, Biermann,
  Creevey, Evans, Eyer, Hutton, Jansen, Jordi, Klioner, Lammers, Lindegren,
  Mignard, Panem, Pourbaix, Randich, Sartoretti, Soubiran, Walton, Arenou,
  {Bailer-Jones}, Bastian, Cropper, Drimmel, Katz, Lattanzi, {van Leeuwen},
  Bakker, Casta{\~n}eda, De~Angeli, Ducourant, Fouesneau, Fr{\'e}mat, Guerra,
  Guerrier, Guiraud, {Jean-Antoine Piccolo}, Masana, Messineo, Mowlavi,
  Nicolas, Nienartowicz, Pailler, Panuzzo, Riclet, Roux, Seabroke, Sordo,
  Tanga, Th{\'e}venin, {Gracia-Abril}, Portell, Teyssier, Altmann, Andrae,
  {Bellas-Velidis}, Benson, Berthier, Blomme, Brugaletta, Burgess, Busso,
  Carry, Cellino, Cheek, Damerdji, Davidson, Delchambre, Dell'Oro,
  {Fern{\'a}ndez-Hern{\'a}ndez}, Galluccio, {Garc{\'i}a-Lario},
  {Garcia-Reinaldos}, {Gonz{\'a}lez-N{\'u}{\~n}ez}, Gosset, Haigron, Halbwachs,
  Hambly, Harrison, Hatzidimitriou, Heiter, Hern{\'a}ndez, Hestroffer, Hodgkin,
  Holl, Jan{\ss}en, {Jevardat de Fombelle}, Jordan, {Krone-Martins}, Lanzafame,
  L{\"o}ffler, Lorca, Manteiga, Marchal, Marrese, Moitinho, Mora, Muinonen,
  Osborne, Pancino, Pauwels, {Recio-Blanco}, Richards, Riello, Rimoldini,
  Robin, Roegiers, Rybizki, Siopis, Smith, Sozzetti, Ulla, Utrilla, {van
  Leeuwen}, {van Reeven}, Abbas, Abreu~Aramburu, Accart, Aerts, Aguado, Ajaj,
  Altavilla, {\'A}lvarez, {\'A}lvarez Cid-Fuentes, Alves, Anderson,
  Anglada~Varela, Antoja, Audard, Baines, Baker, {Balaguer-N{\'u}{\~n}ez},
  Balog, Barache, Barbato, Barros, Barstow, Bartolom{\'e}, Bassilana, Bauchet,
  {Baudesson-Stella}, Becciani, Bellazzini, Bernet, Bertone, Bianchi,
  {Blanco-Cuaresma}, Boch, Bombrun, Bossini, Bouquillon, Bragaglia, Bramante,
  Breedt, Bressan, Brouillet, Bucciarelli, Burlacu, Busonero, Butkevich, Buzzi,
  Caffau, Cancelliere, C{\'a}novas, {Cantat-Gaudin}, Carballo, Carlucci,
  Carnerero, Carrasco, Casamiquela, Castellani, Castro~Sampol, Chaoul, Charlot,
  Chiavassa, Comoretto, Cooper, Cornez, Cowell, Crifo, Crosta, Crowley,
  Dafonte, Dapergolas, David, David, {de Laverny}, De~Luise, De~March,
  De~Ridder, {de Souza}, {de Teodoro}, {de Torres}, {del Peloso}, {del Pozo},
  Delgado, Delisle, Di~Matteo, Diakite, Diener, Distefano, Dolding, Eappachen,
  Enke, Esquej, Fabre, Fabrizio, Faigler, Fedorets, Fernique, Fienga, Figueras,
  Fouron, Fragkoudi, Fraile, Franke, Gai, Garabato, {Garcia-Gutierrez},
  {Garc{\'i}a-Torres}, Gavras, Gerlach, Geyer, Giacobbe, Gilmore, Girona,
  Giuffrida, Gomez, {Gonzalez-Santamaria}, {Gonz{\'a}lez-Vidal}, Granvik,
  {Guti{\'e}rrez-S{\'a}nchez}, Guy, Hauser, Haywood, Hidalgo, Hilger,
  H{\l}adczuk, Hobbs, Holland, Huckle, Jasniewicz, Jonker, Juaristi~Campillo,
  Julbe, Karbevska, Kervella, Khanna, Kochoska, Kontizas, Kordopatis, Korn,
  {Kostrzewa-Rutkowska}, Kruszy{\'n}ska, Lambert, Lanza, Lasne, Le~Campion,
  Le~Fustec, Lebreton, Lebzelter, Leccia, Leclerc, {Lecoeur-Taibi}, Liao,
  Licata, Lindstr{\o}m, Lister, Livanou, Lobel, Madrero~Pardo, Managau, Mann,
  Marchant, Marconi, Marcos~Santos, Marinoni, Marocco, Marshall, Martin~Polo,
  {Mart{\'i}n-Fleitas}, Masip, Massari, {Mastrobuono-Battisti}, Mazeh, Messina,
  Michalik, Millar, Mints, Molina, Molinaro, Moln{\'a}r, Montegriffo,
  Morbidelli, Morel, Morris, Mulone, Munoz, Murphy, Musella, Noval,
  Ord{\'e}novic, Orr{\`u}, Osinde, Pagani, Pagano, Palaversa, Palicio, Panahi,
  Pawlak, Pe{\~n}alosa~Esteller, Penttil{\"a}, Piersimoni, Pineau, Plachy,
  Plum, Poggio, Poretti, Poujoulet, Pr{\v s}a, Pulone, Racero, Ragaini, Rainer,
  Raiteri, Rambaux, Ramos, {Ramos-Lerate}, Re~Fiorentin, Regibo, Reyl{\'e},
  Riva, Rixon, Robichon, Robin, Roelens, Rohrbasser, Rowell, Royer, Rybicki,
  Sadowski, Sagrist{\`a}~Sell{\'e}s, Sahlmann, Salgado, Salguero, Samaras,
  Gimenez, Sanna, Santove{\~n}a, Sarasso, Schultheis, Sciacca, Segol, Segovia,
  S{\'e}gransan, Semeux, Siddiqui, Siebert, Siltala, Slezak, Smart, Solano,
  Solitro, Souami, Souchay, Spagna, Spoto, Steele, Steidelm{\"u}ller,
  Stephenson, S{\"u}veges, Szabados, {Szegedi-Elek}, Taris, Tauran, Taylor,
  Teixeira, Thuillot, Tonello, Torra, Torra, Turon, Unger, Vaillant, {van
  Dillen}, Vanel, Vecchiato, Viala, Vicente, Voutsinas, Weiler, Wevers,
  Wyrzykowski, Yoldas, Yvard, Zhao, Zorec, Zucker, Zurbach, \&
  Zwitter}]{gaiacollaboration2021GaiaEarly}
{Gaia Collaboration}, Luri, X., Chemin, L., {et~al.} 2021, A\&A, 649, A7

\bibitem[{Harris {et~al.}(2020)Harris, Millman, {van der Walt}, Gommers,
  Virtanen, Cournapeau, Wieser, Taylor, Berg, Smith, Kern, Picus, Hoyer, {van
  Kerkwijk}, Brett, Haldane, {del R{\'i}o}, Wiebe, Peterson,
  {G{\'e}rard-Marchant}, Sheppard, Reddy, Weckesser, Abbasi, Gohlke, \&
  Oliphant}]{harris2020ArrayProgramming}
Harris, C.~R., Millman, K.~J., {van der Walt}, S.~J., {et~al.} 2020, Nat, 585,
  357

\bibitem[{Hatzidimitriou {et~al.}(2005)Hatzidimitriou, Stanimirovic,
  Maragoudaki, {Staveley-Smith}, Dapergolas, \&
  Bratsolis}]{hatzidimitriou2005PropertiesHI}
Hatzidimitriou, D., Stanimirovic, S., Maragoudaki, F., {et~al.} 2005, MNRAS,
  360, 1171

\bibitem[{Heiles \& Troland(2003)}]{heiles2003MillenniumArecibo}
Heiles, C. \& Troland, T.~H. 2003, ApJ, 145, 329

\bibitem[{Heitsch {et~al.}(2022)Heitsch, Marchal, {Miville-Desch{\^e}nes},
  Shull, \& Fox}]{heitsch2022MassMorphing}
Heitsch, F., Marchal, A., {Miville-Desch{\^e}nes}, M.~A., Shull, J.~M., \& Fox,
  A.~J. 2022, MNRAS, 509, 4515

\bibitem[{Hunter(2007)}]{hunter2007Matplotlib2D}
Hunter, J.~D. 2007, CiSE, 9, 90

\bibitem[{Inutsuka {et~al.}(2015)Inutsuka, Inoue, Iwasaki, \&
  Hosokawa}]{inutsuka2015FormationDestruction}
Inutsuka, S.-i., Inoue, T., Iwasaki, K., \& Hosokawa, T. 2015, A\&A, 580, A49

\bibitem[{Jameson {et~al.}(2019)Jameson, {McClure-Griffiths}, Liu, Dickey,
  {Staveley-Smith}, Stanimirovi{\'c}, Dempsey, Dawson, D{\'e}nes, Bolatto, Li,
  \& Wong}]{jameson2019ATCASurvey}
Jameson, K.~E., {McClure-Griffiths}, N.~M., Liu, B., {et~al.} 2019, ApJ, 244, 7

\bibitem[{Kalberla {et~al.}(2016)Kalberla, Kerp, Haud, Winkel, Ben~Bekhti,
  Fl{\"o}er, \& Lenz}]{kalberla2016ColdMilky}
Kalberla, P. M.~W., Kerp, J., Haud, U., {et~al.} 2016, ApJ, 821, 117

\bibitem[{Konz {et~al.}(2002)Konz, Br{\"u}ns, \&
  Birk}]{konz2002DynamicalEvolution}
Konz, C., Br{\"u}ns, C., \& Birk, G.~T. 2002, A\&A, 391, 713

\bibitem[{Maia {et~al.}(2014)Maia, Piatti, \&
  Santos}]{maia2014MassDistribution}
Maia, F. F.~S., Piatti, A.~E., \& Santos, J. F.~C. 2014, MNRAS, 437, 2005

\bibitem[{Marchal {et~al.}(2021)Marchal, Martin, \&
  Gong}]{marchal2021ResolvingFormation}
Marchal, A., Martin, P.~G., \& Gong, M. 2021, ApJ, 921, 11

\bibitem[{Marchal {et~al.}(2024)Marchal, Martin, {Miville-Desch{\^e}nes},
  {McClure-Griffiths}, Lynn, Bracco, \& Vujeva}]{marchal2024MappingLower}
Marchal, A., Martin, P.~G., {Miville-Desch{\^e}nes}, M.-A., {et~al.} 2024, ApJ,
  961, 161

\bibitem[{Marchal {et~al.}(2019)Marchal, {Miville-Desch{\^e}nes}, Orieux, Gac,
  Soussen, Lesot, {d'Allonnes}, \& Salom{\'e}}]{marchal2019ROHSARegularized}
Marchal, A., {Miville-Desch{\^e}nes}, M.-A., Orieux, F., {et~al.} 2019, A\&A,
  626, A101

\bibitem[{{Mart{\'i}nez-Delgado} {et~al.}(2019){Mart{\'i}nez-Delgado}, Vivas,
  Grebel, Gallart, Pieres, Bell, Zivick, Lemasle, Clifton~Johnson,
  {Carballo-Bello}, No{\"e}l, Cioni, Choi, Besla, Schmidt, Zaritsky, Gruendl,
  Seibert, Nidever, Monteagudo, Monelli, Hubl, {van der Marel}, Ballesteros,
  Stringfellow, Walker, Blum, Bell, Conn, Olsen, Martin, Chu, Inno, Boer,
  Kallivayalil, De~Leo, Beletsky, Neyer, \&
  Mu{\~n}oz}]{martinez-delgado2019NatureShell}
{Mart{\'i}nez-Delgado}, D., Vivas, A.~K., Grebel, E.~K., {et~al.} 2019, A\&A,
  631, A98

\bibitem[{{McClure-Griffiths} {et~al.}(2018){McClure-Griffiths}, D{\'e}nes,
  Dickey, {Stanimirovi{\'c}}, ~, {Staveley-Smith}, Jameson, Di~Teodoro,
  Allison, Collier, Chippendale, Franzen, G{\"u}rkan, Heald, Hotan, Kleiner,
  {Lee-Waddell}, McConnell, Popping, Rhee, Riseley, Voronkov, \&
  Whiting}]{mcclure-griffiths2018ColdGas}
{McClure-Griffiths}, N.~M., D{\'e}nes, H., Dickey, J.~M., {et~al.} 2018, Nat.
  Astron, 2, 901

\bibitem[{{McClure-Griffiths} {et~al.}(2006){McClure-Griffiths}, Dickey,
  Gaensler, Green, \& Haverkorn}]{mcclure-griffiths2006MagneticallyDominated}
{McClure-Griffiths}, N.~M., Dickey, J.~M., Gaensler, B.~M., Green, A.~J., \&
  Haverkorn, M. 2006, ApJ, 652, 1339

\bibitem[{{McClure-Griffiths} {et~al.}(2015){McClure-Griffiths}, Stanimirovic,
  Murray, Li, Dickey, {Vazquez-Semadeni}, Peek, Putman, Clark,
  {Miville-Deschenes}, {Bland-Hawthorn}, \&
  {Staveley-Smith}}]{mcclure-griffiths2015GalacticMagellanic}
{McClure-Griffiths}, N.~M., Stanimirovic, S., Murray, C., {et~al.} 2015, in
  Advancing {{Astrophysics}} with the {{Square Kilometre Array}} ({{AASKA14}}),
  130

\bibitem[{{McClure-Griffiths} {et~al.}(2023){McClure-Griffiths},
  Stanimirovi{\'c}, \& Rybarczyk}]{mcclure-griffiths2023AtomicHydrogen}
{McClure-Griffiths}, N.~M., Stanimirovi{\'c}, S., \& Rybarczyk, D.~R. 2023,
  ARA\&A, 61, 19

\bibitem[{McCray \& Kafatos(1987)}]{mccray1987SupershellsPropagating}
McCray, R. \& Kafatos, M. 1987, ApJ, 317, 190

\bibitem[{Muraveva {et~al.}(2018)Muraveva, Subramanian, Clementini, Cioni,
  Palmer, {van Loon}, Moretti, {de Grijs}, Molinaro, Ripepi, Marconi, Emerson,
  \& Ivanov}]{muraveva2018VMCSurvey}
Muraveva, T., Subramanian, S., Clementini, G., {et~al.} 2018, MNRAS, 473, 3131

\bibitem[{Murray {et~al.}(2024)Murray, Hasselquist, Peek, Lindberg, Almeida,
  Choi, Craig, D{\'e}nes, Dickey, Di~Teodoro, Federrath, Gerrard, Gibson,
  Leahy, Lee, Lynn, Ma, Marchal, {McClure-Griffiths}, Nidever, Nguyen, Pingel,
  Tarantino, Uscanga, \& {van Loon}}]{murray2024GalacticEclipse}
Murray, C.~E., Hasselquist, S., Peek, J. E.~G., {et~al.} 2024, ApJ, 962, 120

\bibitem[{Noon {et~al.}(2023)Noon, Krumholz, Di~Teodoro, {McClure-Griffiths},
  Lockman, \& Armillotta}]{noon2023DirectObservations}
Noon, K.~A., Krumholz, M.~R., Di~Teodoro, E.~M., {et~al.} 2023, MNRAS, 524,
  1258

\bibitem[{Normandeau {et~al.}(1996)Normandeau, Taylor, \&
  Dewdney}]{normandeau1996GalacticChimney}
Normandeau, M., Taylor, A.~R., \& Dewdney, P.~E. 1996, Nat, 380, 687

\bibitem[{Pidopryhora {et~al.}(2007)Pidopryhora, Lockman, \&
  Shields}]{pidopryhora2007OphiuchusSuperbubble}
Pidopryhora, Y., Lockman, F.~J., \& Shields, J.~C. 2007, ApJ, 656, 928

\bibitem[{Pingel {et~al.}(2022)Pingel, Dempsey, {McClure-Griffiths}, Dickey,
  Jameson, Arce, Anglada, {Bland-Hawthorn}, Breen, {Buckland-Willis}, Clark,
  Dawson, D{\'e}nes, Di~Teodoro, For, Foster, G{\'o}mez, Imai, Joncas, Kim,
  Lee, Lynn, Leahy, Ma, Marchal, McConnell, {Miville-Desch{\`e}nes}, Moss,
  Murray, Nidever, Peek, Stanimirovi{\'c}, {Staveley-Smith}, {Tepper-Garcia},
  Tremblay, Uscanga, {van Loon}, {V{\'a}zquez-Semadeni}, Allison, Anderson,
  Ball, Bell, Bock, Bunton, Cooray, Cornwell, Koribalski, Gupta, Hayman,
  {Harvey-Smith}, {Lee-Waddell}, Ng, Phillips, Voronkov, Westmeier, \&
  Whiting}]{pingel2022GASKAPHIPilot}
Pingel, N.~M., Dempsey, J., {McClure-Griffiths}, N.~M., {et~al.} 2022, PASA,
  39, e005

\bibitem[{Quilis \& Moore(2001)}]{quilis2001WhereAre}
Quilis, V. \& Moore, B. 2001, ApJ, 555, L95

\bibitem[{Roberts(1975)}]{roberts1975RadioObservations}
Roberts, M.~S. 1975, Radio {{Observations}} of {{Neutral Hydrogen}} in
  {{Galaxies}} (University of Chicago Press), 309

\bibitem[{Rolleston {et~al.}(2002)Rolleston, Trundle, \&
  Dufton}]{rolleston2002PresentdayChemical}
Rolleston, W. R.~J., Trundle, C., \& Dufton, P.~L. 2002, A\&A, 396, 53

\bibitem[{Russell \& Dopita(1992)}]{russell1992AbundancesHeavy}
Russell, S.~C. \& Dopita, M.~A. 1992, ApJ, 384, 508

\bibitem[{{Staveley-Smith} {et~al.}(1997){Staveley-Smith}, Sault,
  Hatzidimitriou, Kesteven, \& McConnell}]{staveley-smith1997HIAperture}
{Staveley-Smith}, L., Sault, R.~J., Hatzidimitriou, D., Kesteven, M.~J., \&
  McConnell, D. 1997, MNRAS, 289, 225

\bibitem[{Taank {et~al.}(2022)Taank, Marchal, Martin, \&
  Vujeva}]{taank2022MappingThermal}
Taank, M., Marchal, A., Martin, P.~G., \& Vujeva, L. 2022, ApJ, 937, 81

\bibitem[{Vujeva {et~al.}(2023)Vujeva, Marchal, Martin, \&
  Taank}]{vujeva2023MappingMultiphase}
Vujeva, L., Marchal, A., Martin, P.~G., \& Taank, M. 2023, ApJ, 951, 120

\bibitem[{Wakker(2004)}]{wakker2004HVCIVC}
Wakker, B.~P. 2004, in High {{Velocity Clouds}}, Vol. 312, 25

\bibitem[{Winkler {et~al.}(2015)Winkler, Smith, Points, \& {MCELS
  Team}}]{winkler2015InterstellarMedium}
Winkler, P.~F., Smith, R.~C., Points, S.~D., \& {MCELS Team}. 2015, in Fifty
  {{Years}} of {{Wide Field Studies}} in the {{Southern Hemisphere}}:
  {{Resolved Stellar Populations}} of the {{Galactic Bulge}} and {{Magellanic
  Clouds}}, Vol. 491, 343

\bibitem[{Wolfire {et~al.}(1995)Wolfire, Hollenbach, McKee, Tielens, \&
  Bakes}]{wolfire1995NeutralAtomic}
Wolfire, M.~G., Hollenbach, D., McKee, C.~F., Tielens, A. G. G.~M., \& Bakes,
  E. L.~O. 1995, ApJ, 443, 152

\bibitem[{Wolfire {et~al.}(2003)Wolfire, McKee, Hollenbach, \&
  Tielens}]{wolfire2003NeutralAtomic}
Wolfire, M.~G., McKee, C.~F., Hollenbach, D., \& Tielens, A. G. G.~M. 2003,
  ApJ, 587, 278

\bibitem[{Zucker \& Chen(2018)}]{zucker2018RadFilPython}
Zucker, C. \& Chen, H. H.-H. 2018, ApJ, 864, 152

\end{thebibliography}

\begin{appendix}
\section{Model fitting statistics}
\label{Asec:bestfit}
To ensure that there was no signal left unfit in addition to visually inspecting the reduced chi-squared map (outlined in Section \ref{subsubsec:fitting}, we also looked at the distribution of the residual values across the cloud subcubes. For each subcube we calculate the residuals ($R(x,y,v)$) as below:
\begin{equation}
    R(x,y,v) = Data(x,y,v) - \Sigma_{n=1}^N A_n\exp\left(\frac{-\left(v-\mu_n\right)^2}{2\sigma_n^2}\right),
\end{equation}
where $A$, $\mu$, and $\sigma$ are the fitted components (amplitude, central velocity and dispersion) of $n$ number of Gaussian components.

From this we plot the distribution of all values in the resulting residual cube and calculate the skewness ($\gamma$) (Equation \ref{eq:skew}). 
\begin{equation}
\label{eq:skew}
    \gamma = \frac{\Sigma_i^{N_R}\left(R_i-\bar R\right)^3}{\left(N_R-1\right)\sigma_R^3},
\end{equation}
where $N_R$ is the number of residual values in the subcube, $R$ is the residual value, $\bar R$ is the mean residual value of the subcube, and $\sigma_R$ is the standard deviation of the residuals in the subcube.

\begin{figure}[ht]
    \centering
    \includegraphics[width=0.5\textwidth]{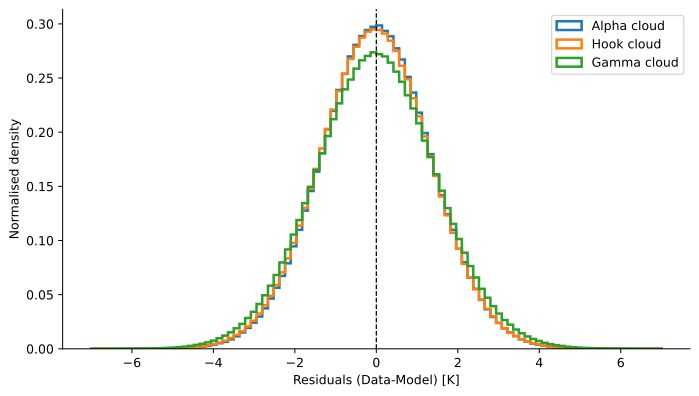}
    \caption{Normalised distribution of the residual values of brightness temperature for all three cloud subcubes. The black dotted line indicates the expected centre at 0.}
    \label{fig:skewness}
\end{figure}

The distributions of the residuals for all three fields are shown in Figure \ref{fig:skewness}. All of the distributions have a calculated skewness of $0-0.01$ which means they are normally distributed, so we do not see any signal in the residual cubes.

\end{appendix}

\end{document}